\documentclass[sn-mathphys-num]{sn-jnl}

\usepackage{graphicx}%
\usepackage{multirow}%
\usepackage{amsmath,amssymb,amsfonts}%
\usepackage{slashed}
\usepackage{amsthm}%
\usepackage{mathrsfs}%
\usepackage[title]{appendix}%
\usepackage{xcolor}%
\usepackage{textcomp}%
\usepackage{manyfoot}%
\usepackage{booktabs}%
\usepackage{algorithm}%
\usepackage{algorithmicx}%
\usepackage{algpseudocode}%
\usepackage{listings}%
\usepackage{MnSymbol}
\theoremstyle{thmstyleone}%
\theoremstyle{thmstyletwo}%

\theoremstyle{thmstylethree}%

\begin{document}
\title[Article Title]{\bf{Effective Lagrangian of a fermion field in a nontrivial topology under magnetic effects}}
\author*[1,2]{\fnm{{Emerson B. S.}} \sur{{Corrêa}}}\email{ecorreae@ufpa.br}

\author*[3]{\fnm{Michelli S. R. Sarges}}
\author*[4]{\fnm{Airton H. G. Rodrigues}}
\author*[5]{\fnm{Carlos A. Bahia}}
\affil*[1]{\orgdiv{Faculdade de Física}, \orgname{Universidade Federal do Pará - UFPA}, %
\postcode{66075-110},
\city{Belém}, \state{PA}, \country{Brazil}}

\affil*[2]{\orgdiv{Programa de Pós-Graduação em Física}, \orgname{Universidade Federal do Sul e Sudeste do Par\'a - Unifesspa},
\postcode{68505-080},
\city{Marabá}, \state{PA}, \country{Brazil}}
\affil*[3]{\orgdiv{Independent researcher}, \orgname{Bel\'em, PA, Brazil}}
\affil*[4]{\orgdiv{Faculdade de Engenharia de Materiais}, \orgname{Universidade Federal do Par\'a, 66075-110, Ananindeua, PA, Brazil}}
\affil*[5]{\orgdiv{Faculdade de Engenharia Elétrica}, \orgname{Universidade Federal do Sul e Sudeste do Par\'a, 68501-970, Marab\'a, PA, Brazil}}




\abstract{

We have investigated a fermionic system from the perspective of an effective quantum field theory defined on a nontrivial topology in the presence of an external magnetic field. Using the proper-time representation, we obtained one-loop expressions for the corresponding effective Lagrangian, taking into account all Landau levels in the propagator. We also computed the significant boundary-induced contributions to the system’s magnetization. To verify the reliability of our results, we examined the limit of zero temperature and infinite spatial extent, which correctly reproduces the celebrated Schwinger result.

}

\keywords{Charged Dirac Field ; Effective Lagrangian ; Toroidal topology; Magnetic effects.}



\maketitle


%
\section{Introduction}


Effective Lagrangians are useful and powerful tools for describing the behaviour of physical systems, even though they provide only a simplified representation of a more complex underlying theory. In general, effective theories offer a simpler theoretical framework for analysing the low-energy sector of a given theory~\cite{Reuter}.

Effective Lagrangians have played a role in the study of strong interactions even before the formulation of quantum chromodynamics (QCD). As it is understood today, the Nambu–Jona-Lasinio (NJL) Lagrangian describes an effective interaction between quarks in the absence of explicit gluonic degrees of freedom~\cite{NJL,NJL1}. The NJL model effectively reproduces key features of QCD in the low–momentum-transfer regime, yielding several notable results, including the inverse magnetic catalysis (IMC) phenomenon~\cite{IMC55,EPJAEmerson2023}. Another important effective Lagrangian for QCD that exhibits spontaneous chiral symmetry breaking—similarly to the NJL model—is the Gross–Neveu model. This model leads to dynamical mass generation for interacting fermions and displays the magnetic catalysis phenomenon~\cite{GN,EmersonEPJC}.

The idea of using effective Lagrangians to incorporate quantum corrections and the effects of external electromagnetic fields into originally classical Lagrangians dates back to early developments in spinor QED. This approach originated in the seminal 1936 paper “Consequences of Dirac’s Theory of the Positrons” by W. Heisenberg and H. Euler~\cite{Heisenberg}. In the same year, V. Weisskopf investigated similar issues in the fermionic sector of QED and extended the analysis to scalar QED in his work “The Electrodynamics of the Vacuum Based on the Quantum Theory of the Electron”~\cite{Weisskopf}. 
Since these pioneering studies—carried out through the calculation of vacuum energy shifts induced by external electromagnetic backgrounds—effective Lagrangians have become a powerful tool across many areas of theoretical physics. They have been employed in analyses of renormalizability and cutoff procedures~\cite{LE-cutoff}, astrophysical environments~\cite{Elmfors3}, lower-dimensional Yang–Mills theories~\cite{Smilga}, non-Abelian background fields~\cite{Dunne}, laser physics~\cite{Laser}, condensed matter systems~\cite{PRX-NR}, and in the contexts of QED and QCD under strong fields~\cite{Itakura}.

Motivated by this rich and evolving topic, we aim to contribute to the study of one-loop effective Lagrangians for fermionic systems in nontrivial scenarios, where boundary conditions—as well as temperature and density effects—play roles as significant as those of external fields. In Refs.~\cite{ElmforsPRL,Elmfors1}, the authors investigated external magnetic fields at specific intensity regimes, namely the strong- and weak-field limits. However, these works did not incorporate topological corrections, which represents a gap in the current literature. The main purpose of this manuscript is to include topological effects in the effective Lagrangians that describe fermion fields subjected to heating and confinement in a finite region of spacetime, without employing any approximations regarding the strength of the external magnetic field. In other words, we intend to apply quantum field theory (QFT) on a toroidal topology within the functional formalism to analyse a fermionic system permeated by an external magnetic field.  

The rigorous mathematical foundations of QFT on a toroidal topology were established only recently~\cite{Book,PR}. In this framework, thermal effects are introduced using the conventional Matsubara formalism. In this approach, the imaginary-time axis no longer extends along an infinite line but instead forms a circle of circumference $\beta = 1/k_{B}T$, where $k_{B}$ is the Boltzmann constant and $T$ is the temperature of the thermal reservoir. Thus, within the Matsubara formalism, temperature becomes a topological feature of the system, corresponding in this case to a cylindrical topology. By extending this analysis to spatial coordinates, one introduces additional circular identifications in the directions ($x_1,x_2,\cdots$) each with lengths ($L_1,L_2,\cdots$). This generalization is known as the generalized Matsubara formalism, and systems described in this manner are defined on a toroidal topology. Recently, applications of this formalism to effective models of QCD have been discussed in Refs.~\cite{PRD2019,PhysicaA,PRD2022,EPJAEmerson2023,PRDEmerson2023}. 

As mentioned above, we were unable to find in the literature a treatment that incorporates the relevant topological effects into the calculation of effective Lagrangians, which is one of our main motivations for conducting this study. In this work, we apply QFT to a fermionic field defined on a toroidal topology in the presence of an external magnetic field, thereby capturing finite-size effects, density, and temperature without invoking the usual reductions of spatial integrals imposed by the magnetic field in thermal–magnetized analyses~\cite{EmersonIJMPB,Avancini}. In fact, this work follows a natural extension of the very recent paper published in Ref. ~\cite{Correa2026} in which a Bose field was treated in a non-trivial topology under magnetic effects. We hope that our results will help fill this gap in the literature by providing the effective Lagrangian for a fermion field in a genuinely nontrivial topology.

The paper is organized as follows. In Sec. II, we obtain the effective Lagrangian for the Dirac field in a toroidal topology. Using the proper-time representation together with the definition of one of the Jacobi theta functions, we derive a unified expression for the fermionic Lagrangian that simultaneously incorporates magnetic, density, finite-volume, and temperature effects. In Sec. III, we address the renormalization procedure, focusing on how topological effects modify the equations. Our prescription removes the divergent vacuum contribution as well as the purely magnetic one. In Sec. IV, we apply our results to compute the magnetization of the system, examining its dependence on the thermodynamic parameters mentioned earlier. In Sec. V, we discuss the results and present our final remarks. Throughout the manuscript, we use natural units and a four-dimensional Euclidean space.
\section{Effective Lagrangian in a toroidal topology under an external magnetic field}
\subsection{Effective Lagrangian and the propagation function}

Let us consider a relativistic fermion quantum gas whose particles have charge $|e|$ and spin half. In coincidence limit, the effective Lagrangian for this system, embedded in a constant magnetic field along the $z$-axis is given by the expression ($1$) from~\cite{ElmforsPRL}, namely
\begin{eqnarray}
\frac{\partial \mathcal{L}_{\mathrm{eff}}}{\partial m} = \lim_{u\rightarrow u^{\prime}} i\,\mathrm{Tr}\left[\mathbb{S}\left(u,u^{\prime},A^{ext}\right)\right],
\label{1}
\end{eqnarray}
where $u=(\tau,x,y,z)\equiv (\tau,{\bf{r}_{\perp}},z)$, and $\mathbb{S}\left(u,u^{\prime},A^{ext}\right)$ is the Dirac propagator under the external magnetic background pointing in z-direction, $m$ is the particle mass represented by the Dirac field, and $\mathrm{Tr}$ is the trace operation over fermion indices. We use the gauge $A^{ext}_{\mu}=(0,0,xB,0)$. Integrating Eq.~(\ref{1}), we obtain
\begin{eqnarray}
\mathcal{L}_{\mathrm{eff}} = \lim_{u\rightarrow u^{\prime}}i\int\,\mathrm{Tr}\left[\mathbb{S}\left(u,u^{\prime},A^{ext}\right)\right]\,dm + const.
\label{2}
\end{eqnarray}
The integration constant does not depend on the fermion mass. 
We fix it as the Maxwellian term, i.e., $const = B^{2}/2$.

The Dirac propagator (as well as the Bose propagator in Ref.~\cite{Lawrie}) under the magnetic background along the $z$-direction is not invariant under translation in the $xy$-plane. This is because Schwinger phase $\Phi({\bf{r}_{\perp}},{\bf{r}_{\perp}^{\prime}})$. To see that, consider the complete fermion propagator under a magnetic field along z-direction (see e.g., Ref.~\cite{Miransky}),
\begin{eqnarray}
\mathbb{S}\left(u,u^{\prime},A\right) &=& \exp\left[i\,e\,\Phi(\bf{r}_{\perp},\bf{r}_{\perp}^{\prime})\right]\,\int_{0}^{\infty}  dS \,\int \,\frac{d^{4}p}{(2\pi)^{4}} \left\{\frac{}{}i\slashed{p} - im_{0} +\left[\gamma_{x}p_{y}-\gamma_{y}p_{x}\right]\,\tanh{(\omega S)}\right\} \nonumber \\
&\times&\exp\left[ip_{\mu}\left(u-u^{\prime}\right)_{\mu}\right] \,\exp\left\{-S \left[p_{\tau}^{2}+p_{z}^{2}+(p_{x}^{2}+p_{y}^{2})\frac{\tanh{(\omega S)}}{\omega S}+m_{0}^{2}\right]\right\} \nonumber \\
&\times&\left[\mathbb{I}_{4} - \left(\mathbb{I}_{2}\otimes\sigma_{z}\right) \tanh{(\omega S)}\right],
\label{chaves68}
\end{eqnarray}
where $S$ is the so-called proper-time variable, and Schwinger's phase $\Phi({\bf{r}_{\perp}},{\bf{r}_{\perp}^{\prime}})$ is given by
\begin{eqnarray}
\Phi({\bf{r}_{\perp}},{\bf{r}_{\perp}^{\prime}})  =   \frac{B}{2}\left[\left(x+x^{\prime}\right)\left(y-y^{\prime}\right)\right]. 
\label{phase}
\end{eqnarray}
 Notice that in the limit $B \rightarrow 0$, Eq.~(\ref{chaves68}) correctly reduces to the free electron propagator.

The propagator is not translationally invariant in the plane perpendicular to the external field, since Schwinger's phase breaks the translation invariance in ($xy$) plane. Notwithstanding, we can do a gauge transformation to remove the translationally non-invariant part of it.

Initially, let us show that Schwinger's phase is given by integrating the vector potential along a straight line connecting the initial and final points in the ($xy$) plane, i.e.,
\begin{eqnarray}
\Phi(\mathbf{r}_{\perp},\mathbf{r}_{\perp}^{\prime}) = \int_{\mathbf{r}_{\perp}^{\prime}}^{\mathbf{r}_{\perp}}\,\vec{A}(\vec{\xi}\,)\cdot d{\vec{\xi}},
\end{eqnarray}
where
\begin{eqnarray*}
\vec{A}(\vec{\xi}\,)  = \left(0,B\xi_{1},0\right) \,\,\,\Rightarrow\,\,\vec{\nabla}_{\vec{\xi}} \times \vec{A}(\vec{\xi}\,)={B}\hat{z}.
\end{eqnarray*}
Following the Chapter 3 of Ref.~\cite{LivroSch}), the straight line can be parameterized in terms of a parameter $\lambda$:
\begin{eqnarray*}
\vec{\xi}  = (1-\lambda)\,\vec{r}_{\perp}^{\,\prime}+\lambda\,\vec{r}_{\perp},\,\,\,\,\,\Rightarrow\,\,\,\,\, d\vec{\xi} = (\vec{r}_{\perp}-\vec{r}_{\perp}^{\,\prime})\,d\lambda, \,\,\,\,\,\,0\leq\lambda\leq1.
\end{eqnarray*}
Thus,
\begin{eqnarray}
\int_{\mathbf{r}_{\perp}^{\prime}}^{\mathbf{r}_{\perp}}\,\vec{A}(\vec{\xi}\,)\cdot d{\vec{\xi}} &=& \int_{0}^{1}\,B\left[(1-\lambda)\,x^{\prime}+\lambda \,x\right](y-y^{\prime})\,d\lambda \nonumber\\
&=&\frac{B}{2}(x+x^{\prime{}})(y-y^{\prime})\equiv\Phi(\mathbf{r}_{\perp},\mathbf{r}_{\perp}^{\prime}).
\label{6}
\end{eqnarray}

Therefore, the factor responsable for non-invariance under ($xy$) plane in Eq.~(\ref{chaves68}) can be rewritten as  
\begin{eqnarray}
\exp\left[i\,e\,\Phi(\mathbf{r}_{\perp},\mathbf{r}_{\perp}^{\prime})\right]=\exp\left[i\,e\int_{\mathbf{r}_{\perp}^{\prime}}^{\mathbf{r}_{\perp}}\,\vec{A}(\vec{\xi}\,)\cdot d{\vec{\xi}}\,\right].
\label{chaves56}
\end{eqnarray}

It is well known that we have the freedom to choose an arbitrary scalar function $\Lambda(\vec{\xi})$ such that it does not affect the external field. Therefore, we will obtain an equivalent expression for the propagation function according to the prescription
\begin{eqnarray}
 \vec{A} \, \mapsto \, \vec{A} - \vec{\nabla} \, \Lambda \, \Rightarrow \, 
	\begin{array}{lcl}
\mathbb{S}\left(u,u^{\prime},A\right)  \, \mapsto \,	\mathbb{S}\left(u,u^{\prime},A-\partial \Lambda\right).
  	\end{array}
    \label{gauge4}
\end{eqnarray}

A convenient choice for the arbitrary function is $\Lambda(\vec{\xi}\,) = {B(x+x^{\prime})\,\xi_{2}}\,/\,{2}$. Its gradient, when integrated along the straight line, is given by
\begin{eqnarray}
\int_{\mathbf{r}_{\perp}^{\prime}}^{\mathbf{r}_{\perp}}\,\vec{\nabla}\,\Lambda(\vec{\xi}\,)\cdot d{\vec{\xi}} = \frac{B}{2}(x+x^{\prime})(y-y^{\prime}).
\label{55}
\end{eqnarray}

Therefore, under the gauge transformation defined in Eq.~(\ref{gauge4}) and using Eqs.~(\ref{6}) and (\ref{55}), the Green’s function of the fermionic field in the infinite volume in a magnetic background becomes~\cite{Correa-Abdalla}
\begin{eqnarray}
\mathbb{S}\left(u-u^{\prime},\omega\right) &=& \int_{0}^{\infty}  dS \,\int \,\frac{d^{4}p}{(2\pi)^{4}} \left\{\frac{}{}i\slashed{p} - im +\left[\gamma_{x}p_{y}-\gamma_{y}p_{x}\right]\,\tanh{(\omega S)}\right\} \nonumber \\
&\times&\exp\left[ip_{\nu}\left(u-u^{\prime}\right)_{\nu}\right] \,\exp\left\{-S \left[p_{\tau}^{2}+p_{z}^{2}+(p_{x}^{2}+p_{y}^{2})\frac{\tanh{(\omega S)}}{\omega S}+m^{2}\right]\right\} \nonumber \\
&\times&\left[\mathbb{I}_{4} - \left(\mathbb{I}_{2}\otimes\sigma_{z}\right)\tanh{(\omega S)}\right],
\label{chaves69}
\end{eqnarray}
where $\omega \equiv |e|\mathrm{B}$ is the cyclotron frequency. 
\subsection{Treatment of the system in a toroidal topology: temperature effects}
Let us derive the Matsubara formalism expressions within the proper-time representation. We begin with the frequencies associated with the imaginary-time coordinate. In this case, the fermionic nature of the system imposes antiperiodic boundary conditions along the $\tau$-direction over the fermion field $\psi$
\begin{eqnarray}
\psi(\tau+\beta,x_{j}) &=& - \,\, \psi(\tau,x_{j}),\,\,\,\,j=1,2,3.
\end{eqnarray}
Encoding such boundary conditions in the propagator $\mathbb{S}\,(\tau,x_{j},\omega\,)$, implies that
\begin{eqnarray}
\mathbb{S}\,(\tau+\beta,x_{j},\omega) &=& \exp{(i \,\pi\,\alpha_\tau)} \,\mathbb{S}\,(\tau,x_{j},\omega),\,\,\,\,\alpha_\tau \equiv 1,
\end{eqnarray}
which means
\begin{eqnarray}
 \int \,\frac{dp_{\tau}}{2\pi}\,\exp{\left[-p_{\tau}^{2}\,f(S)\right]} & \rightarrow & \frac{1}{\beta }\,\sum_{ n_{\tau}\,=\,-\infty}^{\infty} \,\exp\left\{-\frac{4\pi^{2}\,f(S)}{\beta^{2}}\left[n_{\tau}-\left(i\frac{\mu\beta}{2\pi}-\frac{c_{\tau}}{2}\right)\right]^{2}\right\}, \nonumber\\
p_{\tau} &\rightarrow&  {\omega}_{n_\tau} \equiv \frac{\pi}{\beta}
\left(2  n_{\tau} + c_{\tau} \right)-i\mu, \,\,\,\,\,\,\, n_{\tau} = 0,\pm 1 , \pm 2, \cdots, 
 \label{Mat121}
\end{eqnarray}
where $c_{\tau} = 1$. The chemical potential of the system is denoted by $\mu$, and its temperature by $\beta^{-1}$. The function $f(S)$ represents the contribution of Schwinger’s proper-time. We now make use of the Jacobi identity [see \cite{Elizalde}, Eq. ($4.10$)]
\begin{eqnarray}
\sum_{n_{\tau}\,=\,-\infty}^{+\infty} \,\exp{\left[-a_{\tau}(n_{\tau}-b_\tau)^{2}\right]} &=&\sqrt{\frac{\pi}{a_\tau}}\,\sum_{n_{\tau}\,=\,-\infty}^{+\infty}\,\exp{\left(-\pi^{2}\,n_{\tau}^{2}/a_{\tau}\right)}\,\exp{\left(2\pi i\,b_{\tau}n_{\tau}\right)},
\label{Jacobi}
\end{eqnarray}
with $(a_{\tau},b_{\tau}) \in \mathbb{C}$ such that the real part of $a_{\tau}$ is positive. 

Substituting Eq. (\ref{Jacobi}) into Eq. (\ref{Mat121}) yields
\begin{eqnarray}
 \int \,\frac{dp_{\tau}}{2\pi}\,\exp{\left[-p_{\tau}^{2}\,f(S)\right]} & \rightarrow & \frac{1}{\beta }\,\sqrt{\frac{\pi}{a_\tau}}\,\theta_{3}\left[\pi b_{\tau};\exp{\left(-\pi^{2}/a_\tau\right)}\right], \,\,\mathrm{for} \nonumber \\
 a_{\tau}=\frac{4\pi^{2} f(S)}{\beta^{2}} &,& \,b_{\tau} = \frac{i\beta \mu}{2\pi}-\frac{c_\tau}{2}.
 \label{Mat23}
\end{eqnarray}
where the third Jacobi theta function is defined by~\cite{Whittaker,Optics}
\begin{eqnarray*}
\theta_{3}(u\,;\,\!q) = \sum_{n\,=\,-\infty}^{+\infty}\,q^{n^{2}}\,\exp{\left(2\,n\,i\,u\right)},\,\,\,|q|<1.
\end{eqnarray*}
\subsection{Treatment of the system in a toroidal topology: finite size effects}

We now consider a fermionic system confined to a finite-volume box under a magnetic field. The setup is illustrated in Fig.~1.
\begin{figure}
\centering
\includegraphics[{width=7.50cm}]{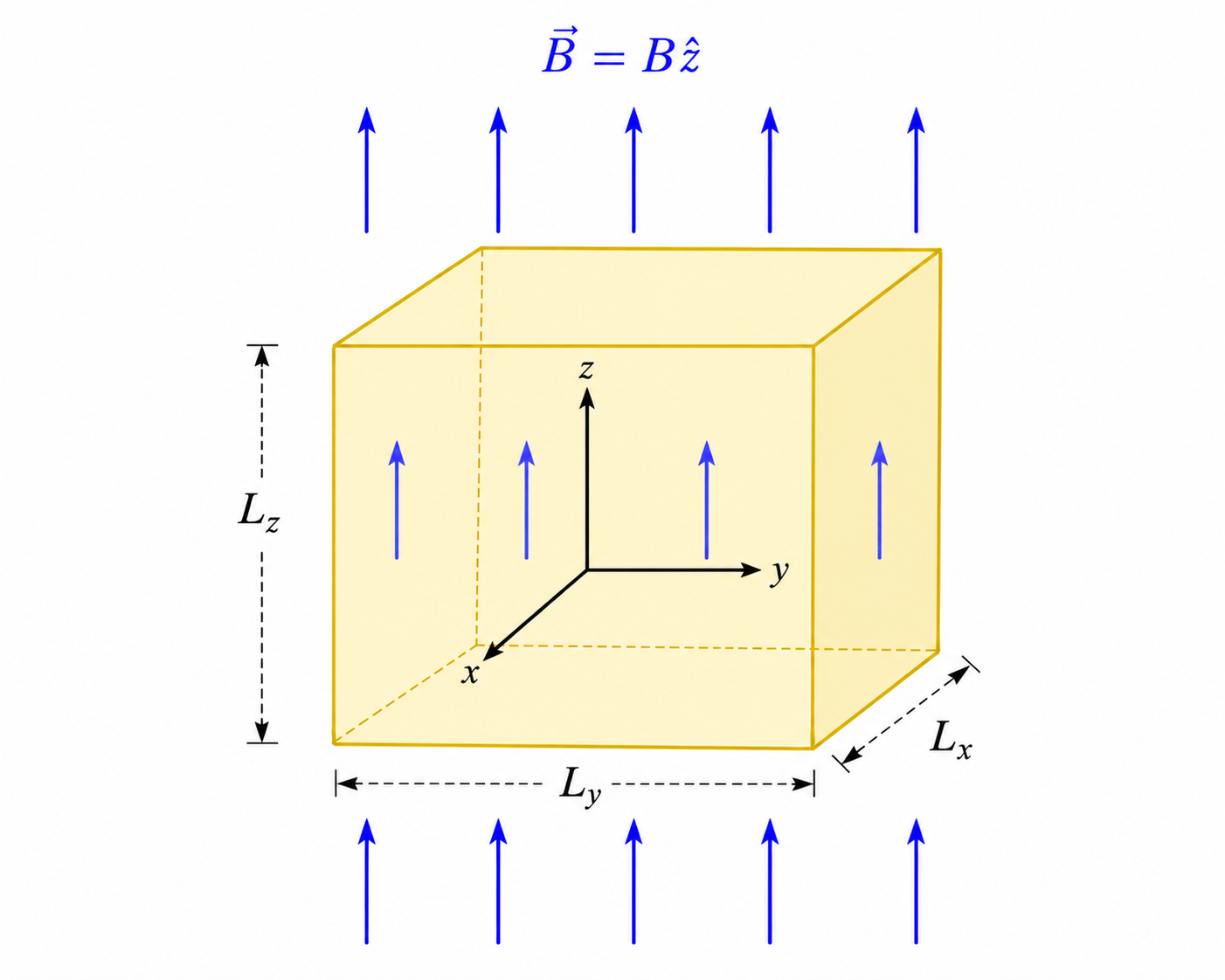}
\caption{~Fermionic field confined within a cubic geometry under a magnetic field along the $z$ direction.}
\end{figure}

The compact manifold $\left(T^{3}_{xyz}\right)$, known as a {\textit{three-torus}}, is obtained by identifying the opposite faces of the cube shown in Fig.~1. Topologically, two of these compactifications are represented in Fig.~2, forming a {\textit{two-torus}}~\cite{Cinthia}.
\begin{figure}
\centering
\includegraphics[{width=11.0cm}]{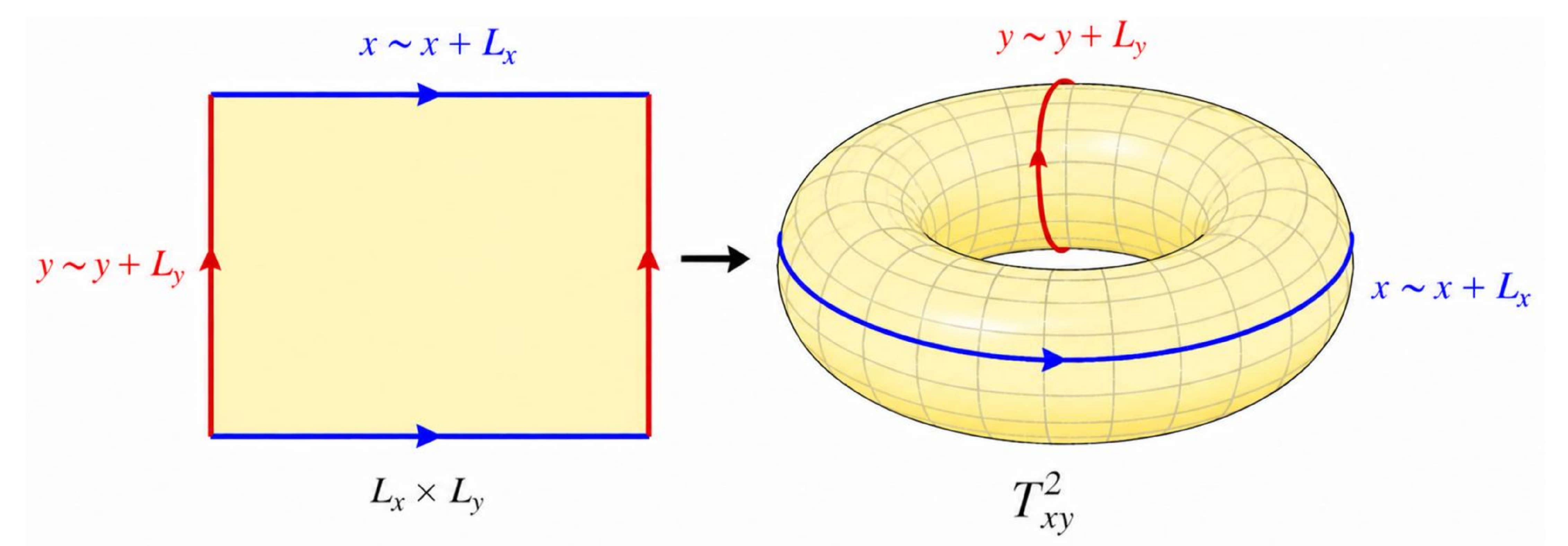}
\caption{~Configuration of the system in the plane orthogonal to the magnetic field, with compactifications in the $x$ and $y$ directions generating a two-torus.}
\end{figure}

The vector potential $\vec{A} = (0,Bx,0)$, which generates the magnetic field along the $z$ direction shown in Fig.~1, must be periodic in the toroidal topology, namely~\cite{Tofit,Accidental}
\begin{eqnarray}
\left\{
\begin{array}{rcl}
\vec{A}(x+L_x,y,z) &=& \vec{A}(x,y,z), \\
\vec{A}(x,y+L_y,z) &=& \vec{A}(x,y,z), \\
\vec{A}(x,y,z+L_z) &=& \vec{A}(x,y,z).
\end{array}
\right.
\label{chave}
\end{eqnarray}

Using the vector potential $\vec{A} = (0,Bx,0)$, the first relation in Eq.~(\ref{chave}) is not satisfied:
\begin{eqnarray*}
\vec{A}(x+L_x,y,z) = B(x+L_x) \hat{y} = \vec{A}(x,y,z) +  BL_z \hat{y} \neq \vec{A}(x,y,z).
\end{eqnarray*}

To satisfy the three-torus boundary conditions on the vector potential defined in Eq.~(\ref{chave}), we perform the gauge transformation
$
\vec{A} \;\mapsto\; \vec{A} - \vec{\nabla} \varphi
$,
where $\varphi$ is a scalar function, and subsequently incorporate its effects on the fermionic field~\cite{Accidental}. Accordingly, we introduce
\begin{eqnarray}
\left\{
\begin{array}{rcl}
\vec{A}(x+L_x,y,z) &=& \vec{A}(x,y,z) -\vec{\nabla} \varphi_{x}, \\
\vec{A}(x,y+L_y,z) &=& \vec{A}(x,y,z)-\vec{\nabla} \varphi_{y}, \\
\vec{A}(x,y,z+L_z) &=& \vec{A}(x,y,z)-\vec{\nabla} \varphi_{z}.
\end{array}
\right.
\label{chave1}
\end{eqnarray}
The functions $\varphi_{i}$ are known as transition functions. Using the expressions in Eq.~(\ref{chave1}) together with the choice of vector potential $\vec{A} = (0,Bx,0)$, we obtain
\begin{eqnarray}
\left\{
\begin{array}{rcl}
\varphi_{x} &=& -B L_{x} y +\theta_{x}, \\
\varphi_{y} &=& \theta_{y},  \\
\varphi_{z} &=& \theta_{z},
\end{array}
\right.
\label{chave2}
\end{eqnarray}
being $\theta_{i}$ parameters that label a family of solutions.

Gauge invariance is implemented for the Dirac field $\psi$ through
\begin{eqnarray}
\left\{
\begin{array}{rcl}
\psi(x+L_x,y,z) &=& \exp[{i e \varphi_{x}}]\,\psi(x,y,z) , \\
\psi(x,y+L_y,z) &=& \exp[{i e \varphi_{y}}]\,\psi(x,y,z), \\
\psi(x,y,z+L_z) &=& \exp[{i e \varphi_{z}}]\,\psi(x,y,z),
\end{array}
\right.
\label{chave3}
\end{eqnarray}
with $e$ denoting the electric charge of the fermion field.

Specifically, on a {\textit{two-torus}} of finite rectangular size 
$L_x \times L_y$, the magnetic flux is quantized~\cite{Barton}. 
To verify this quantization, we perform two successive translations 
using the first and second relations given in Eq.~(\ref{chave3}):
\begin{eqnarray}
\psi(x+L_x,y+L_y,z) &=& \exp\{{i e \varphi_{x}}\}\,\psi(x,y+L_y,z), \nonumber \\
&=&  \exp\left\{i e [-BL_x (y+L_y)+\theta_x]\right\}\, \psi(x,y+L_y,z), \nonumber \\
&=& \exp\left\{i e [-BL_x (y+L_y)+\theta_x + \theta_y]\right\}\, \psi(x,y,z).
\label{a}
\end{eqnarray}

Conversely, the two translations may also be performed in the reverse order by applying the second and first relations presented in Eq.~(\ref{chave3}), i.e.,
\begin{eqnarray}
\psi(x+L_x,y+L_y,z) &=& \exp\{{i e \varphi_{y}}\}\,\psi(x+L_x,y,z), \nonumber \\
&=&  \exp\left\{i e \theta_y\right\}\, \psi(x+L_x,y,z), \nonumber \\
&=& \exp\left\{i e [\theta_y-BL_x y+\theta_x ]\right\}\, \psi(x,y,z).
\label{b}
\end{eqnarray}

From Eq.~(\ref{a}) and Eq.~(\ref{b}), we conclude that they are equivalent only if
\begin{eqnarray}
e[BL_xL_y] \equiv 2\pi \,n \,\Rightarrow\, BL_xL_y = (2\pi/e)\,n,\,\,\,\,\,n\,\in\,\mathbb{Z}.
\end{eqnarray}

As a consequence, the flux $B L_x L_y$ is quantized in units of the {\textit{flux quantum}} $(2\pi/e)$, where $e$ is the charge of the field $\psi$. Therefore, for fixed values of the area $(L_x \,L_y)$, the magnetic field $B$ assumes discrete values inside the box.

Grouping Eq.~(\ref{chave3}), with $\varphi_{j},\,\,j=1,2,3,\,(1 \leftrightarrow x,\,2 \leftrightarrow y,\,3 \leftrightarrow z)$ given by Eq.~(\ref{chave2}), the relations can be rewritten by taking $\theta_{j} \equiv (\pi \alpha_{j})/e$, where $\alpha_j$ is a continuous parameter defined in the interval $0 \leq \alpha_j \leq 1$,
\begin{eqnarray}
\psi(x_j+L_j) &=& \exp[{i \pi \alpha_{j}}]\,\psi(x_j).
\label{prescricao}
\end{eqnarray}

The prescription introduced in Eq.~(\ref{prescricao}) is exact for $j=2$ and $j=3$. In contrast, for $j=1$, this prescription constitutes only an approximation due to the presence of an extra factor $(-BL_x y)$ in $\varphi_1$, as follows from the first equation in Eq.~(\ref{chave2}) together with the first equation in Eq.~(\ref{chave3}). At the boundaries associated with the $y$ coordinate, the prescription given in Eq.~(\ref{prescricao}) becomes exact for $j=1$, specifically at $y=0$ and $y=L_y$.

Accordingly, throughout the paper we adopt the prescription given in Eq.~(\ref{prescricao}), although this spatial prescription may not be exact in the $x$ direction, it remains exact in the other two spatial directions and, for an initial analysis of the model, it provides a valid approximation.

To incorporate topological effects into the propagator, we employ the prescription written in Eq.~(\ref{prescricao}) on the Eq.~(\ref{chaves69}). Consequently, we obtain
\begin{eqnarray}
\mathbb{S} \, (\tau,x_{j}+L_{j},\omega\,) &=& \exp{(i\,\pi\,\alpha_{j})} \, \mathbb{S} \,(\tau,x_{j},\omega\,),\,\,\,\,j=1,2,3,
\end{eqnarray}
where $\alpha_{j} = 0$ denotes periodic boundary conditions, and $\alpha_{j} = 1$
 denotes antiperiodic boundary conditions along the compactified direction $x_j$.
 For values $0 < \alpha_{j} < 1$, the system is subject to mixed (twisted or quasiperiodic) boundary conditions. In this case, the momentum takes discrete values given by
\begin{eqnarray}
p_{j} &\rightarrow & {\omega}_{n_j} \equiv \frac{\pi}{L_{j}}
\left(2  n_{j} \right)-i\mu_{j}, \,\,\,\,\,\,\, n_{j} = 0,\pm 1 , \pm 2, \cdots,\nonumber \\
&\mathrm{where}& \,\,\,\,\,\,\,\,\,\,\,\,\,\,\,\,\,\,\mu_{j}=i\frac{\pi}{L_{j}}\alpha_{j},\,\,\,\,\,\,0\leq \,\alpha_{j}\,\leq 1.
\end{eqnarray}

Therefore, the topological effects in the spatial \(j\)-direction, in the proper-time representation, are incorporated into the fermionic system through the relation
\begin{eqnarray}
 \int \,\frac{dp_{j}}{2\pi}\,\exp{\left[-p_{j}^{2}\,f(S)\right]} & \rightarrow & \frac{1}{L_{j} }\,\sqrt{\frac{\pi}{a_j}}\,\theta_{3}\left[\pi b_{j};\exp{\left(-\pi^{2}/a_j\right)}\right], \,\,\,\mathrm{for} \nonumber \\ \nonumber \\
 a_{j} = \frac{4\pi^{2}\,f(S)}{L_{j}^{2}}& , & b_j= i \frac{\mu_{j}L_{j}}{2\pi} = - \frac{\alpha_{j}}{2}.
 \label{Mat117}
\end{eqnarray}
\subsection{Effective Lagrangian on a four-torus}

Through Eqs.~(\ref{2}) and (\ref{chaves69}), besides that performing the trace operation over Dirac indices and subsequently integrating with respect to the mass $m$, we arrive at the effective Lagrangian
\begin{eqnarray}
\mathcal{L}^{}_{\mathrm{eff}} &=& \frac{B^2}{2}-2\,\int_{0}^{\infty}  \frac{dS}{S} \,\int \,\frac{d^{4}p}{(2\pi)^{4}}\, \exp\left\{-S \left[p_{\tau}^{2}+p_{z}^{2}+(p_{x}^{2}+p_{y}^{2})\frac{\tanh{(\omega S)}}{\omega S}+m^{2}\right]\right\}. \nonumber \\
\label{EffeF}
\end{eqnarray}
Eq.~(\ref{EffeF}) describes the effective interactions of fermions with charge $|e|$ and spin $1/2$ under a constant magnetic background $\mathrm{\bf{B}}$ in infinite space and at zero temperature.

Applying Eqs. (\ref{Mat117}) and (\ref{Mat23}) to the fermionic gas in a magnetic field as defined in Eq. (\ref{EffeF}), and noting that $f(S)=S$ for the $\tau$ and $z$ coordinates, while $f(S) = \tanh{(\omega S)}/\omega$ for the $x$ and $y$ coordinates, we obtain
\begin{eqnarray}
\mathcal{L}^{}_{\mathrm{eff}} &=& \frac{B^2}{2}-\,\frac{1}{8\pi^{2}}\,\int_{0}^{\infty}  \frac{dS}{S^{3}} \,(\omega S) \, \coth{(\omega S)}\, \exp{\left(-S \, m^{2}\right)} \nonumber \\
&\times& \theta_{3}\left[\left(\frac{i\mu\beta}{2}-\frac{\pi}{2}\right);\exp{\left(-\frac{\beta^{2}}{4\,S}\right)}\right]\,\theta_{3}\left[\left(-\frac{\pi \alpha_{z}}{2}\right);\exp{\left(-\frac{L_{z}^{2}}{4\,S}\right)}\right] \nonumber \\
&\times&  \theta_{3}\left[\left(-\frac{\pi \alpha_{x}}{2}\right);\exp{\left(-\frac{\omega L_{x}^{2} \coth{(\omega S)}}{4}\right)}\right]\,\theta_{3}\left[\left(-\frac{\pi \alpha_{y}}{2}\right);\exp{\left(-\frac{\omega L_{y}^{2} \coth{(\omega S)}}{4}\right)}\right]. \nonumber \\
\label{EffeFFx}
\end{eqnarray}

Eq.~(\ref{EffeFFx}) encapsulates the magnetic, topological, thermal, and chemical-potential contributions to the fermionic system, and constitutes the principal result of this work, notwithstanding the presence of ultraviolet divergences. In the following section, we address these divergences by applying the additive renormalization procedure.
\vspace{0.25cm}
\section{Renormalization}

From Eq.~(\ref{EffeFFx}), the series expansion around 
$S \approx 0$ takes the form
 \begin{eqnarray*}
\frac{(\omega S) \coth{(\omega S)}}{S^3} &\approx& \frac{1}{S^{3}} + \frac{\omega^{2}}{3S}+ \mathcal{O}(\omega^{4}S).
 \end{eqnarray*}
 Thus, our prescription for renormalizing the effective Lagrangian at one-loop approximation in Eq.~(\ref{EffeFFx}) will be
 \begin{eqnarray}
\mathcal{L}^{(1)}_{\mathrm{eff}}(\beta,\mu,{\bf{L}},{\bf{\alpha}}) &=& -\,\frac{1}{8\pi^{2}}\,\int_{0}^{\infty}  \frac{dS}{S^{3}} \, \exp{\left(-S \, m^{2}\right)} \nonumber \\
&\times& \left\{\theta_{3}\left[\left(\frac{i\mu\beta}{2}-\frac{\pi}{2}\right);\exp{\left(-\frac{\beta^{2}}{4\,S}\right)}\right]\,\theta_{3}\left[\left(-\frac{\pi \alpha_{z}}{2}\right);\exp{\left(-\frac{L_{z}^{2}}{4\,S}\right)}\right] \right.\nonumber \\
&\times&  \theta_{3}\left[\left(-\frac{\pi \alpha_{x}}{2}\right);\exp{\left(-\frac{\omega L_{x}^{2} \coth{(\omega S)}}{4}\right)}\right]\,\theta_{3}\left[\left(-\frac{\pi \alpha_{y}}{2}\right);\exp{\left(-\frac{\omega L_{y}^{2} \coth{(\omega S)}}{4}\right)}\right] \nonumber \\
&\times&\left. (\omega S) \, \coth{(\omega S)}-1-\frac{(\omega S)^{2}}{3}\right\}.
\label{EffeFFxx}
\end{eqnarray}
Fig.~$3$ illustrates the diagrammatic expansion corresponding to the effective Lagrangians at one-loop order computed in Eq.~(\ref{EffeFFxx}). In the next section, we use this equation to study the magnetization of a fermion gas confined in a finite-volume box and subjected to an external magnetic field at fixed temperature and density.
\begin{figure}
\centering
\includegraphics[{width=13.0cm}]{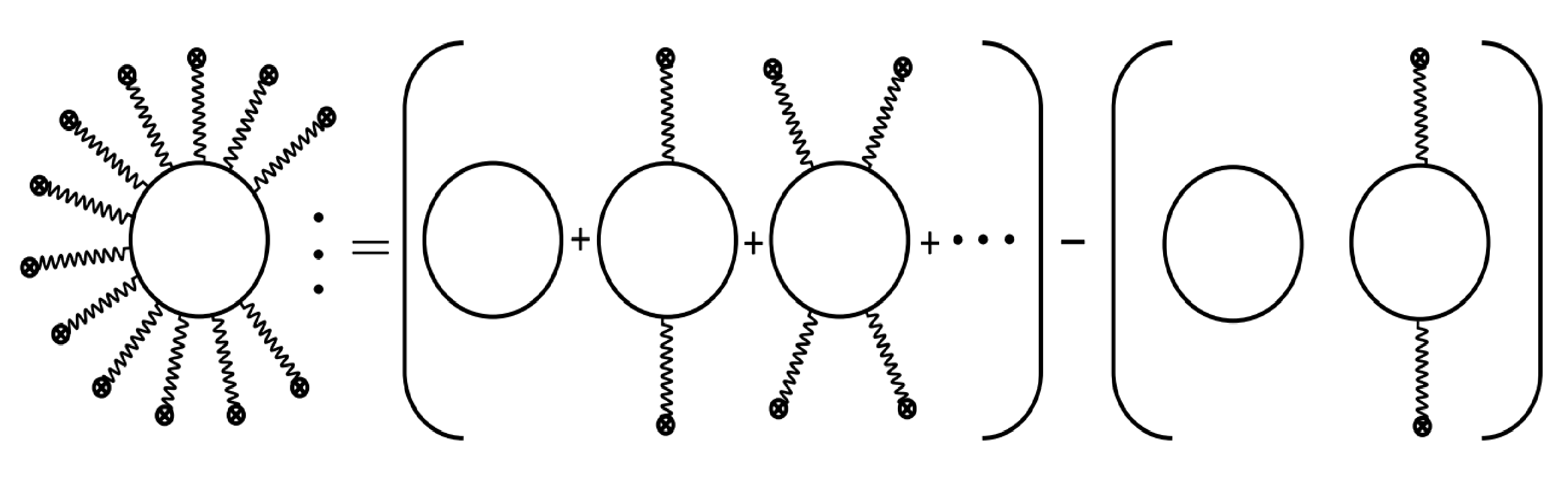} 
\caption{~Diagrammatic expansion corresponding to the effective Lagrangian at one-loop order computed in Eq.~(\ref{EffeFFxx}). Note the two subtracted diagrams in the figure: the vacuum diagram and the one with the $\omega^{2}$ contribution.}
\label{Fig0}
\end{figure}

As a final remark in this section, let us derive the effective Lagrangian at one-loop order in the particular case of infinite volume (bulk form) and zero temperature. To do this, we take the limit $(\beta V) \rightarrow \infty $ in Eq.~(\ref{EffeFFxx}) and use the fact that the Jacobi theta functions approach unity in this limit. The resulting expression is
 \begin{eqnarray}
\lim_{(\beta V) \,\rightarrow \,\infty}\mathcal{L}^{(1)}_{\mathrm{eff}}(\beta,\mu,{\bf{L}},{\bf{\alpha}}) &=& -\,\frac{1}{8\pi^{2}}\,\int_{0}^{\infty}  \frac{dS}{S^{3}} \, \exp{\left(-S \, m^{2}\right)} \nonumber \\
&\times&\left[ (\omega S) \, \coth{(\omega S)}-1-\frac{(\omega S)^{2}}{3}\right],
\label{EffeFFxxx}
\end{eqnarray}

Eq.~(\ref{EffeFFxxx}) corresponds to Schwinger’s formula for the fermion effective Lagrangian at zero temperature and in infinite space [see Eq.~(3.47) of Ref.~\cite{Schwinger}], up to a logarithmically divergent factor that Schwinger absorbed into the Maxwell Lagrangian through a rescaling of the external field and the electric charge.
\section{Magnetization and discussion of the results}

 From now on, we proceed with the application of the effective Lagrangian at one-loop derived in Eq.~(\ref{EffeFFxx}) to describe the topological effects on the magnetization of the system. By functional differentiation~\cite{RevEndrodi},
\begin{eqnarray}
{M}_{B} = \frac{\partial \mathcal{L}^{(1)}_{\mathrm{eff}}}{\partial \omega},
\label{magnetizacao}
\end{eqnarray}
we analyze the response of the systems to changes in the external parameters. For this purpose, we introduce a set of dimensionless quantities, each normalized by the corresponding fermion mass. The reduced variables are: reduced proper time $s$, reduced magnetic field $\delta$, reduced chemical potential $\gamma$, reduced temperature $t$, reduced length $\ell$, reduced volume $v$, and reduced magnetization $m_B$~\cite{PLA}
 \begin{eqnarray}
s \equiv m^{2} \, S \,\,;\,\,\delta \equiv \frac{\omega}{m^{2}}\,\,;\,\,\gamma \equiv \frac{\mu}{m}\,\,;\,\,t \equiv \frac{T}{m}\,\,;\,\, \ell \equiv L \, m \,\,;\,\,v \equiv V \,m^{3} \,\,;\,\,m_{B} \equiv \frac{M_{B}}{m^{2}}.
 \end{eqnarray} 

In Fig.~4, we present the dimensionless magnetization as a function of the reduced temperature at fixed volume. We observe that the magnetization increases as the temperature rises. This behavior is further amplified for stronger external magnetic fields, regardless of the boundary conditions employed. We also note that the magnetization exhibits only a weak dependence on the chemical potential. In all simulated scenarios, the magnetization develops a plateau at a specific value of $\delta$.
\begin{figure}
\centering
\includegraphics[{width=6.45cm}]{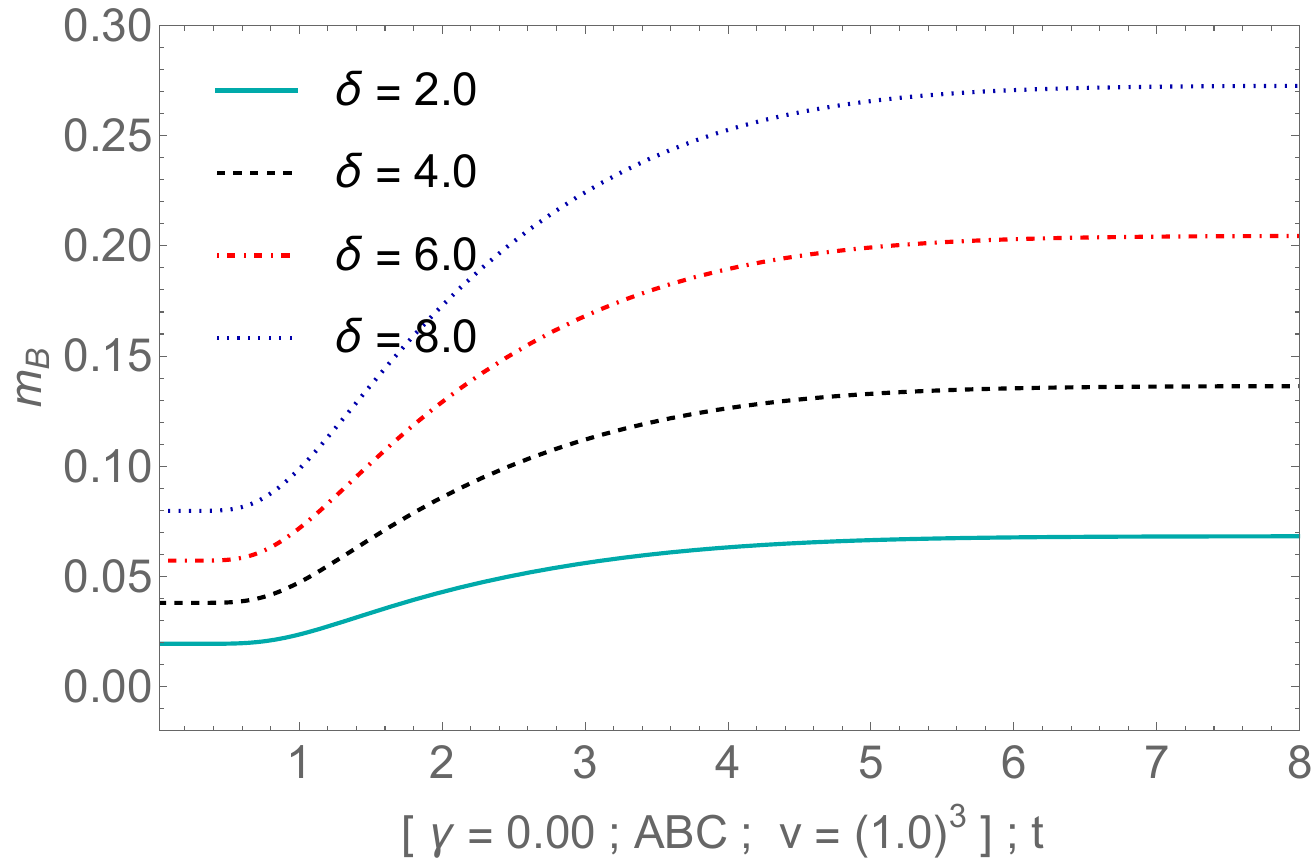}
\includegraphics[{width=6.45cm}]{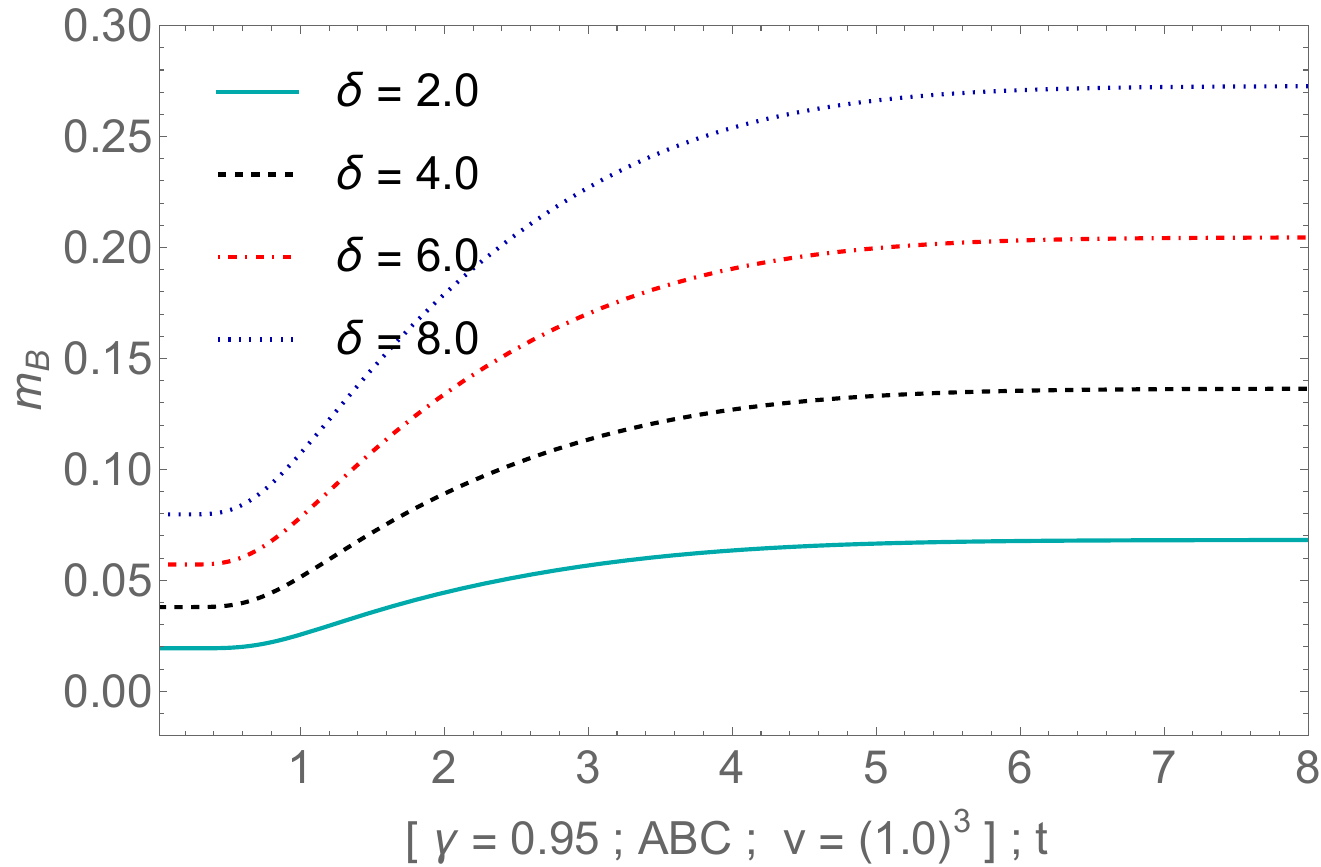} \\
\includegraphics[{width=6.45cm}]{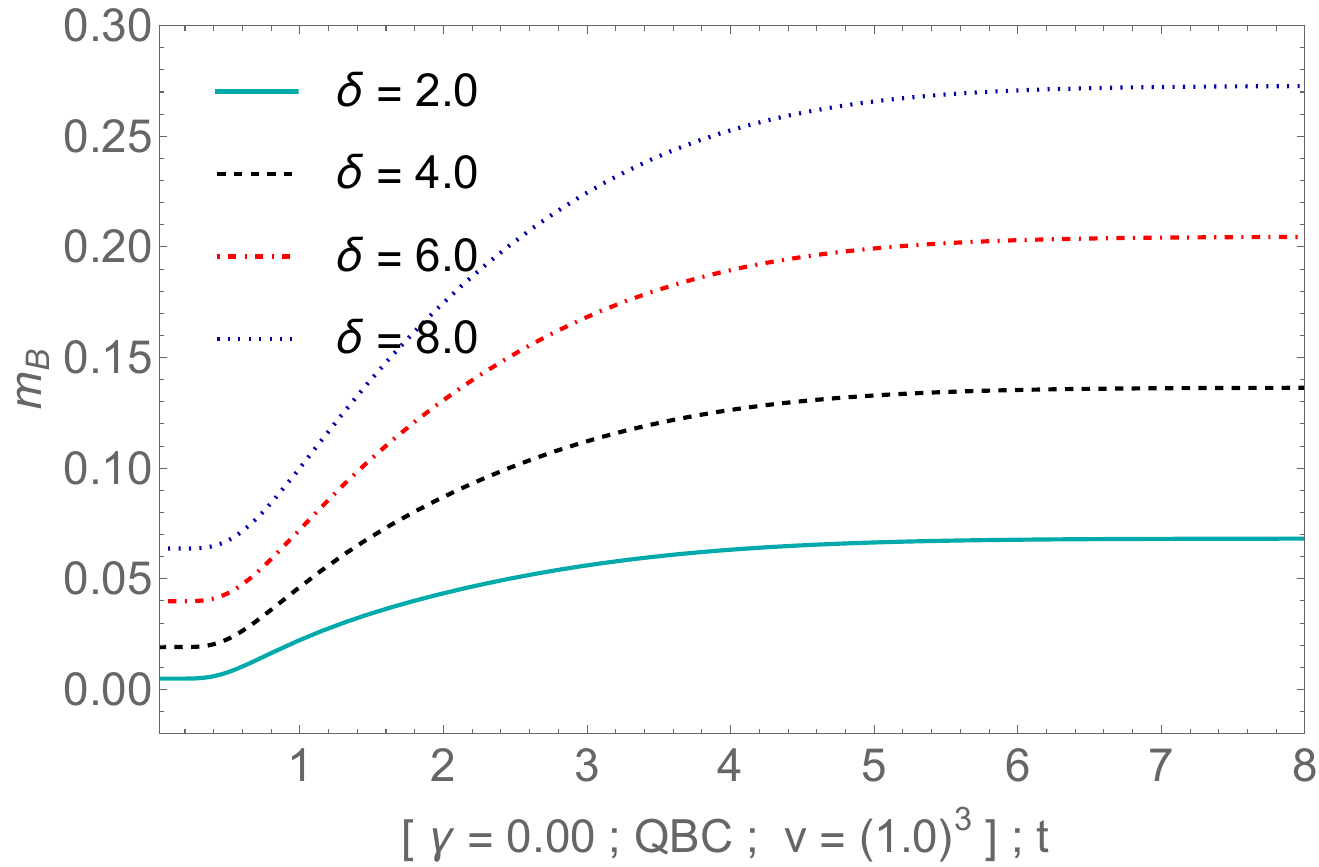}
\includegraphics[{width=6.45cm}]{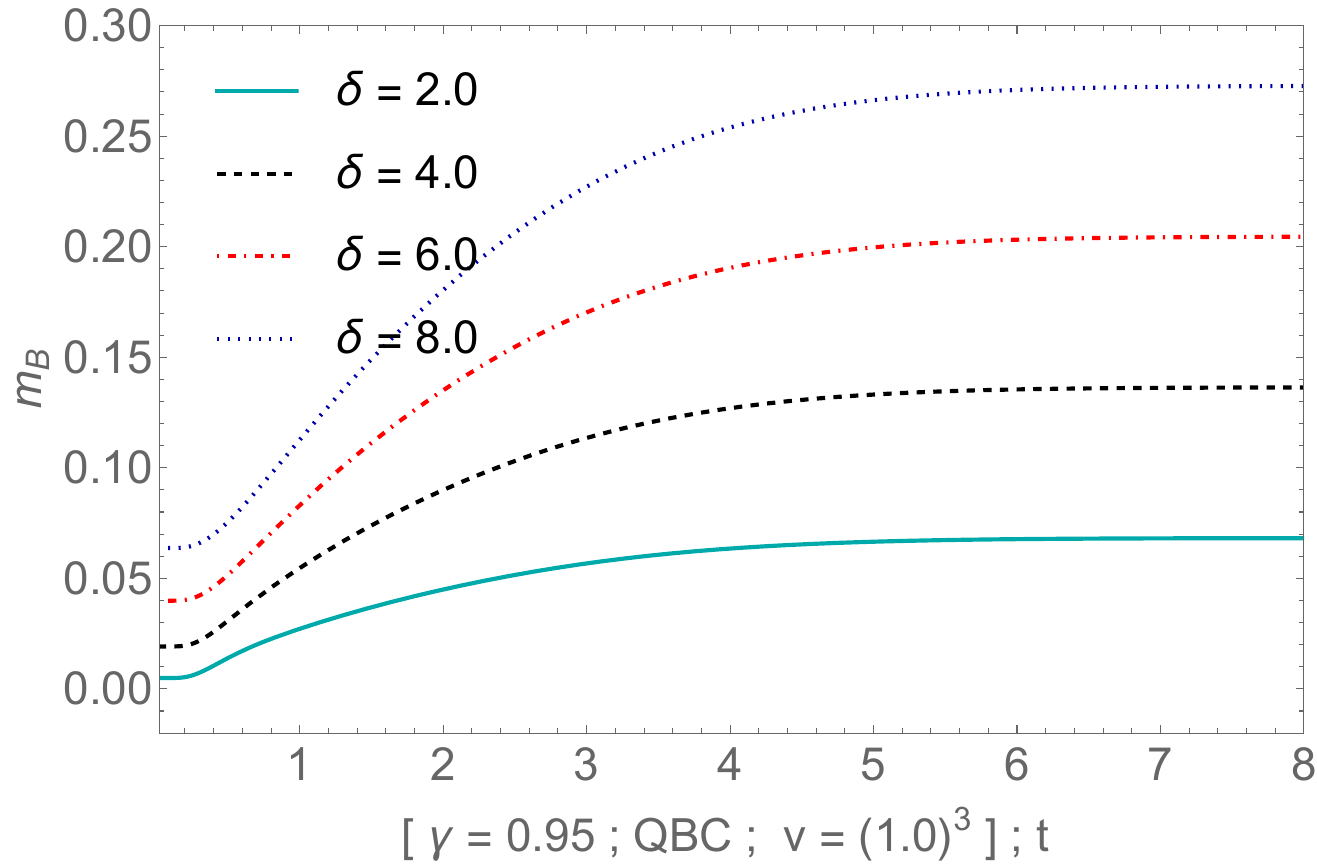}\\
\includegraphics[{width=6.45cm}]{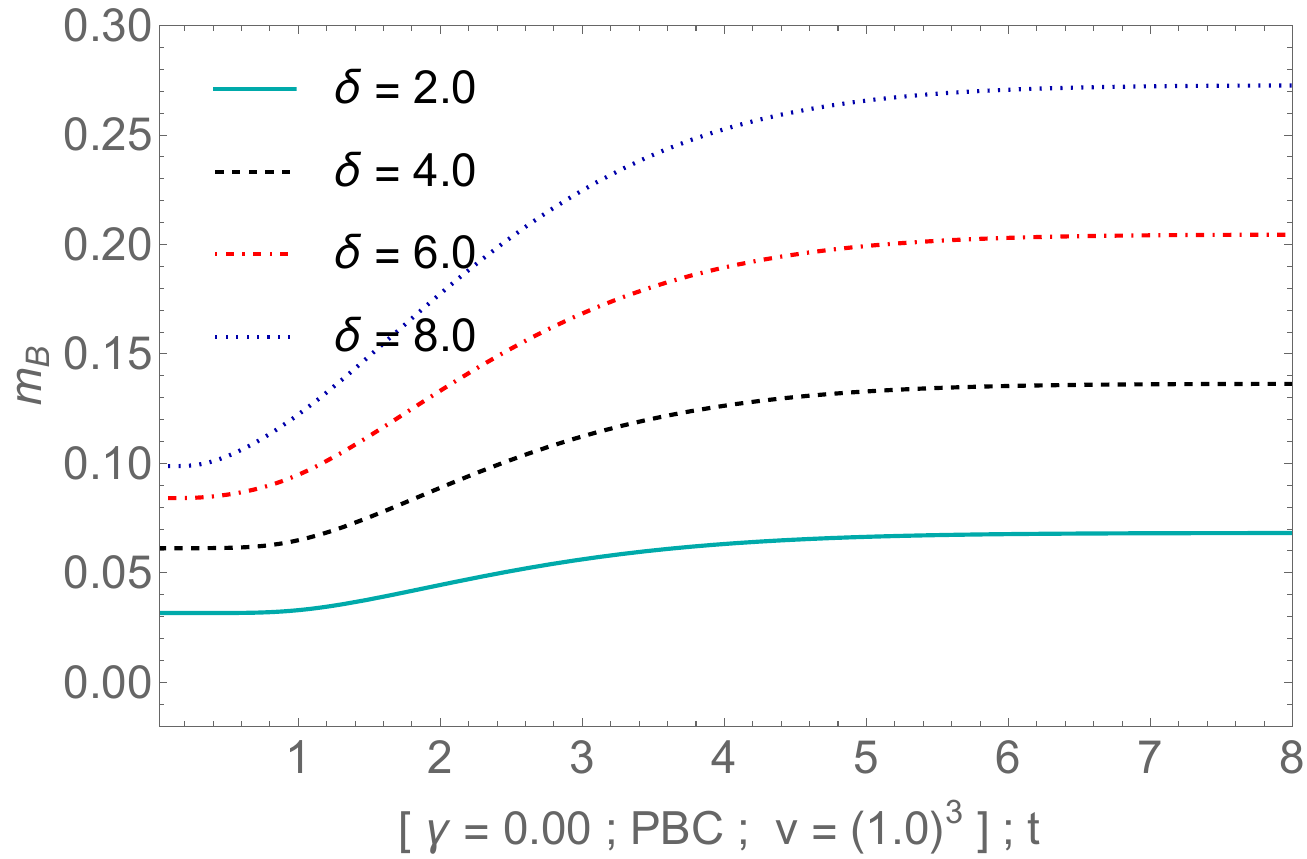}
\includegraphics[{width=6.45cm}]{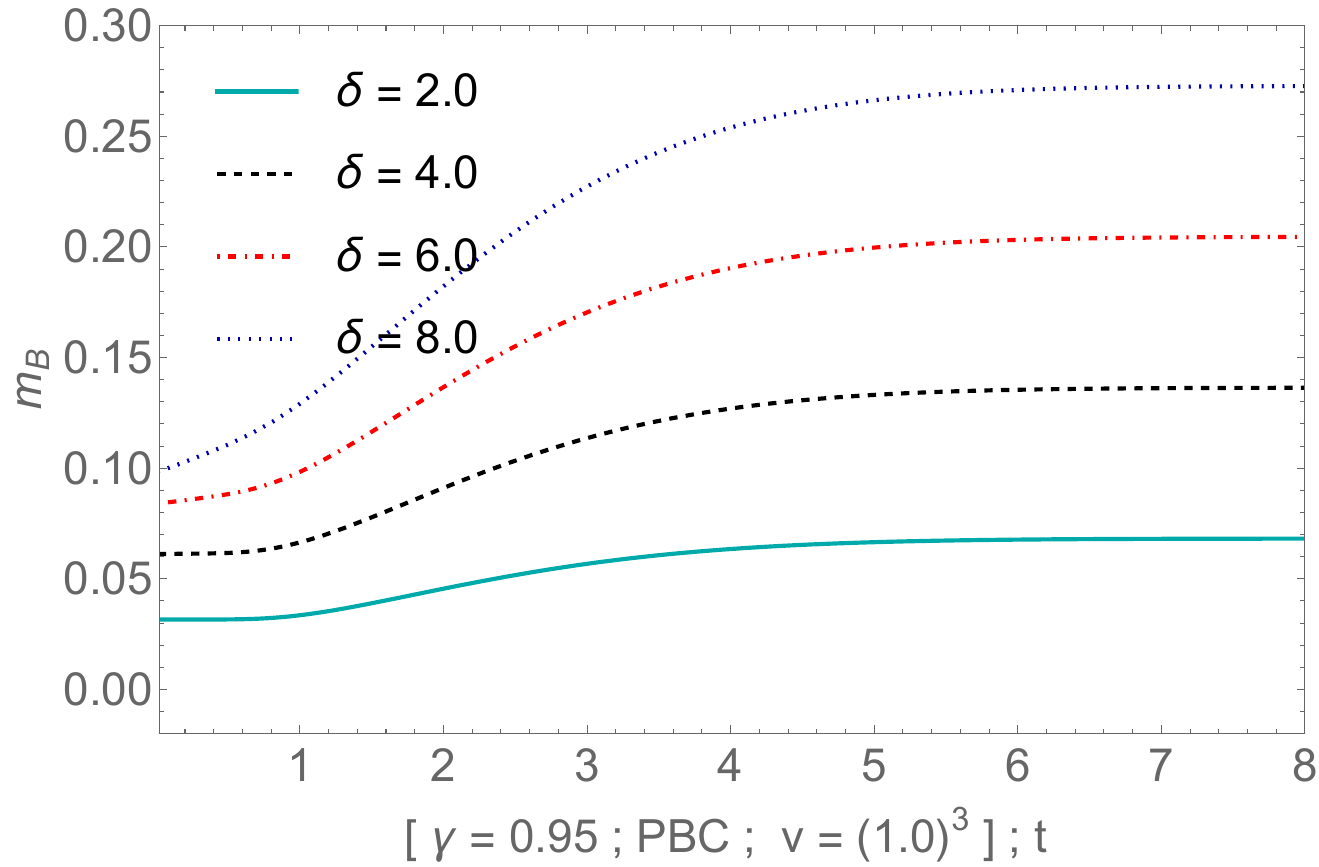} \\
\caption{~Dimensionless magnetization of a Dirac gas as a function of the reduced temperature for fixed external magnetic fields and reduced volume. The left panels correspond to a neutral gas, while the right panels represent a system with a charge imbalance. Boundary conditions are as follows: ABC (top panels), QBC (middle panels), and PBC (bottom panels).}
\label{Fig2}
\end{figure}

In Fig.~5, we show the behavior of the dimensionless magnetization as a function of system volume. At a fixed temperature, $m_B$ decreases as the volume increases, although plateaus are again observed. Each boundary condition exhibits two plateaus: the first occurs at small volumes, while the second (larger) appears at higher volumes. The amplitude of these plateaus depends on the type of boundary condition employed.
\begin{figure}
\centering
\includegraphics[{width=6.45cm}]{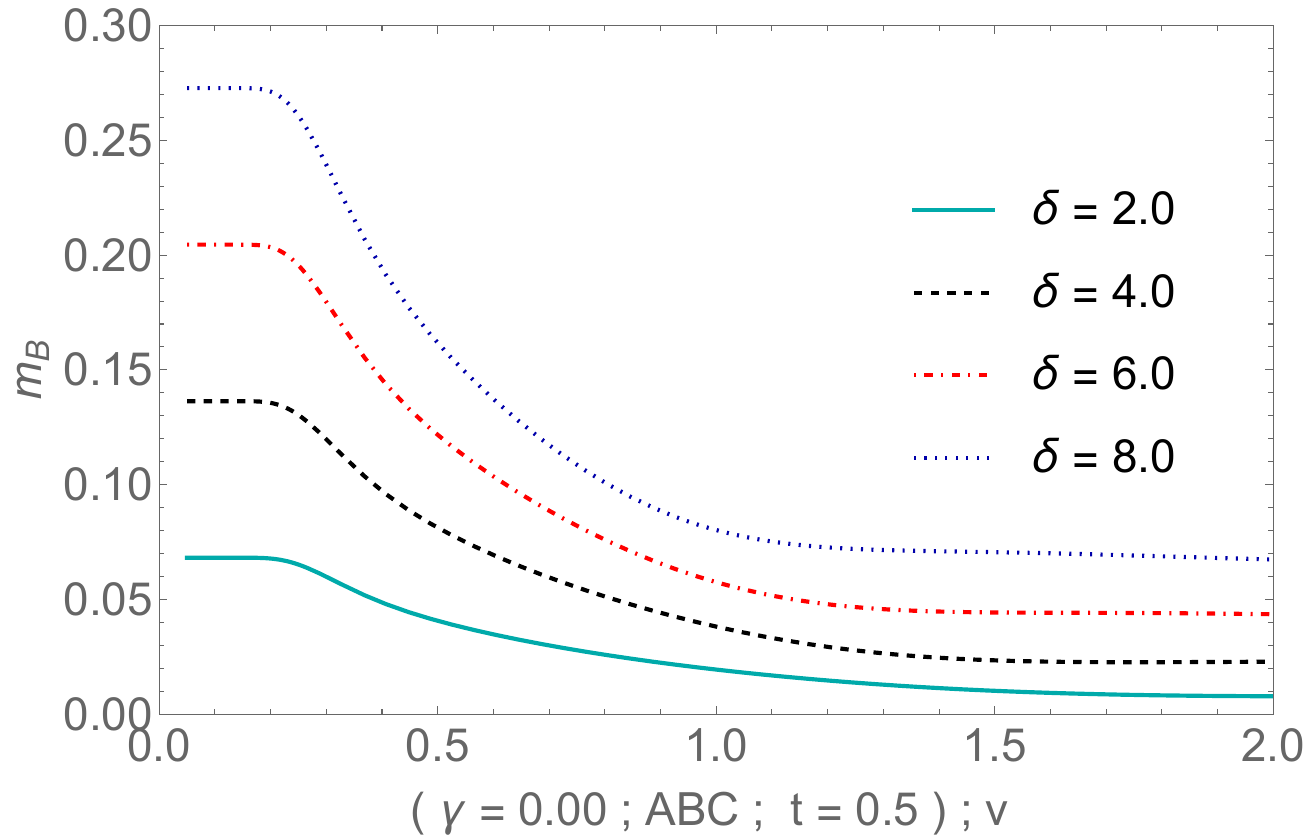} 
\includegraphics[{width=6.45cm}]{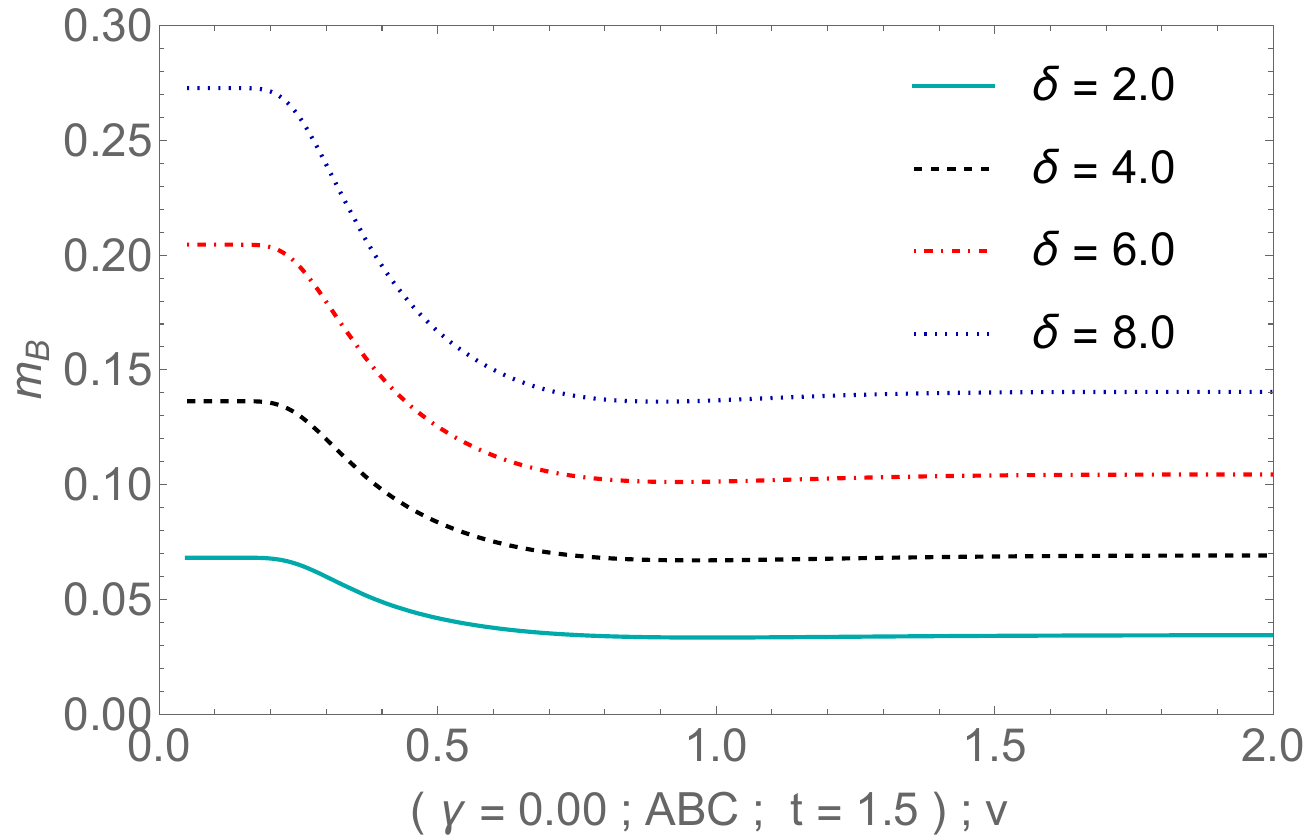} \\
\includegraphics[{width=6.45cm}]{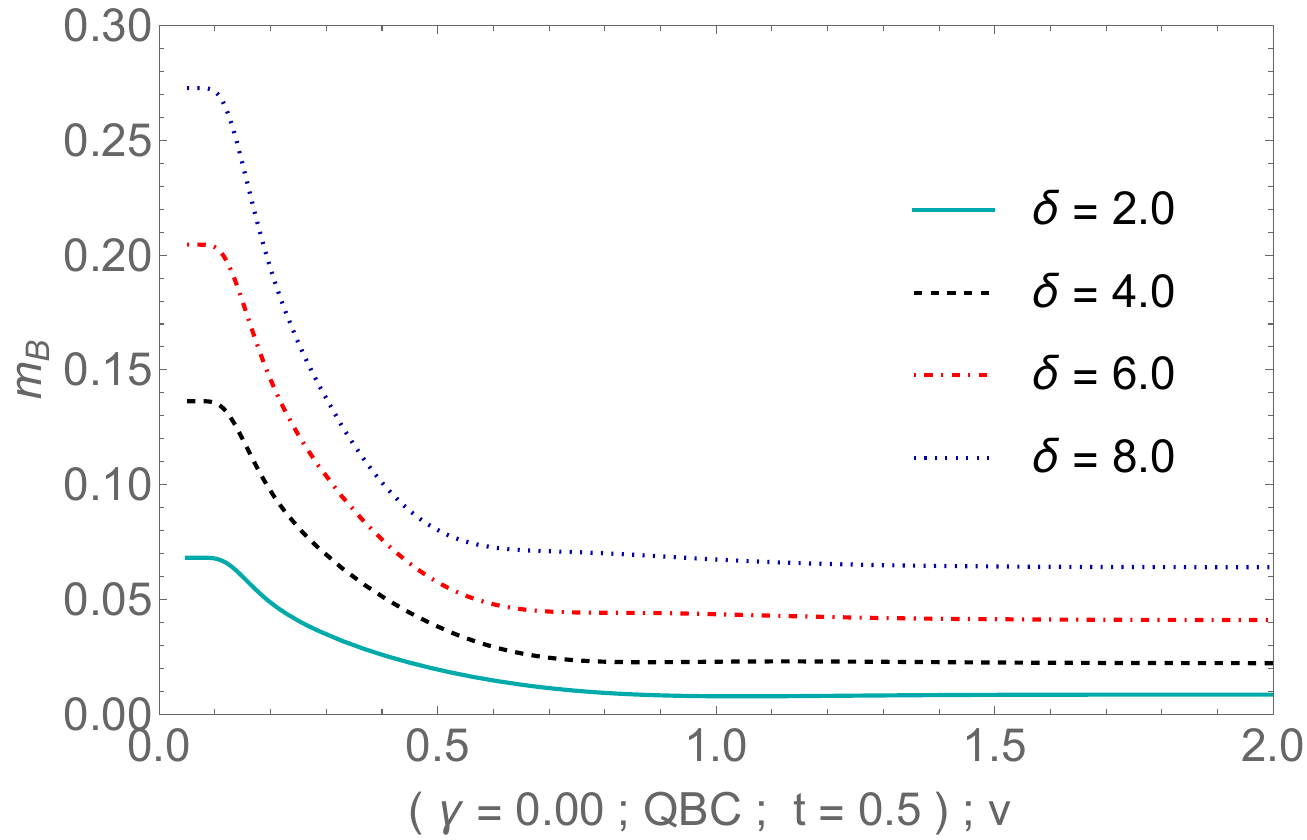}
\includegraphics[{width=6.45cm}]{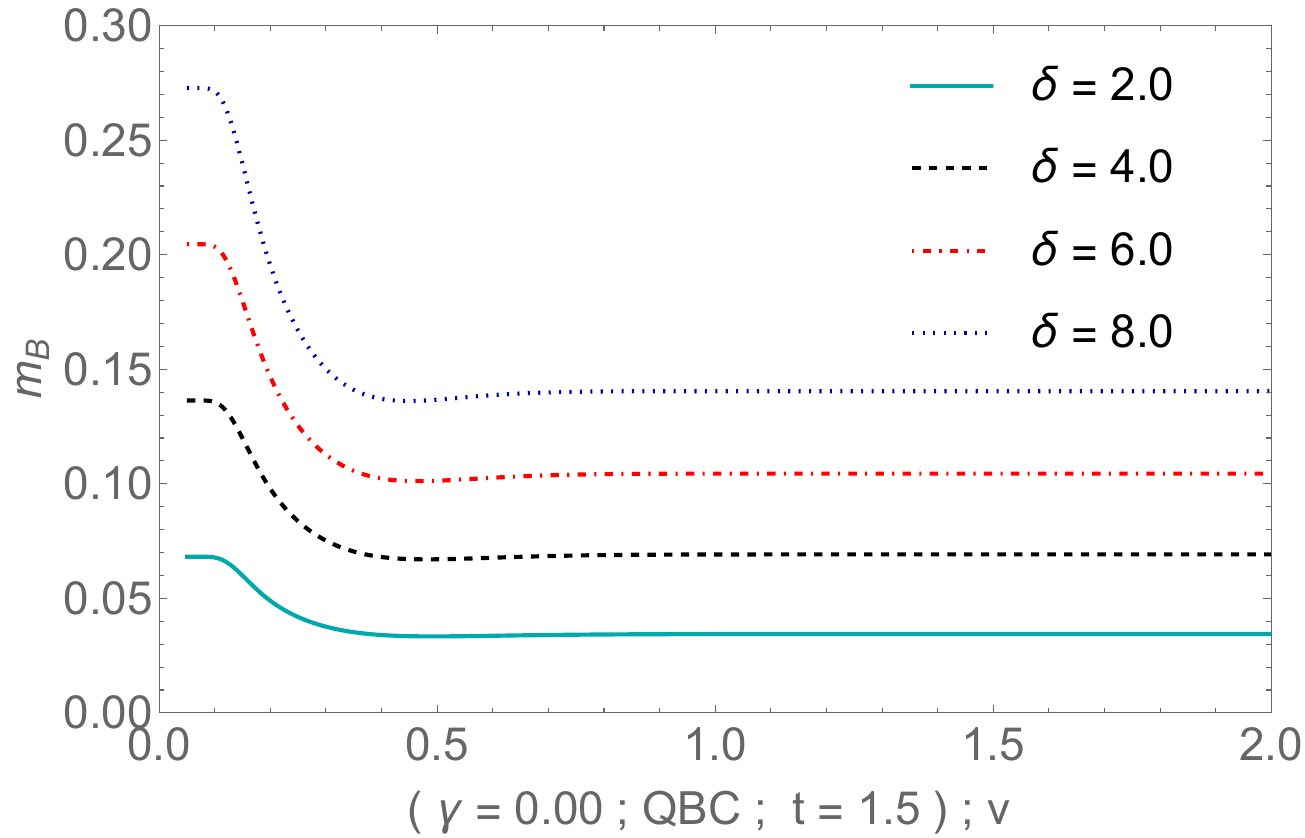}\\
\includegraphics[{width=6.45cm}]{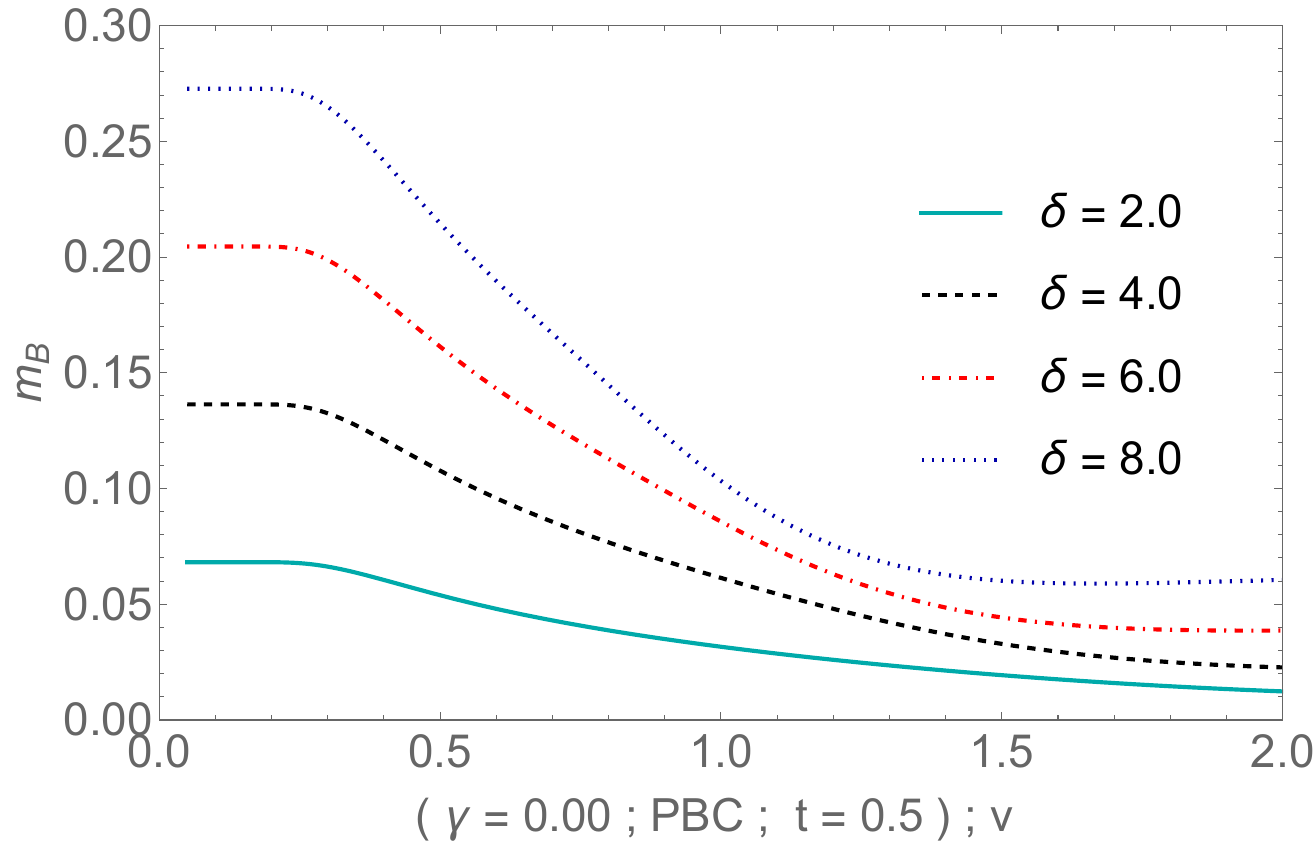}
\includegraphics[{width=6.45cm}]{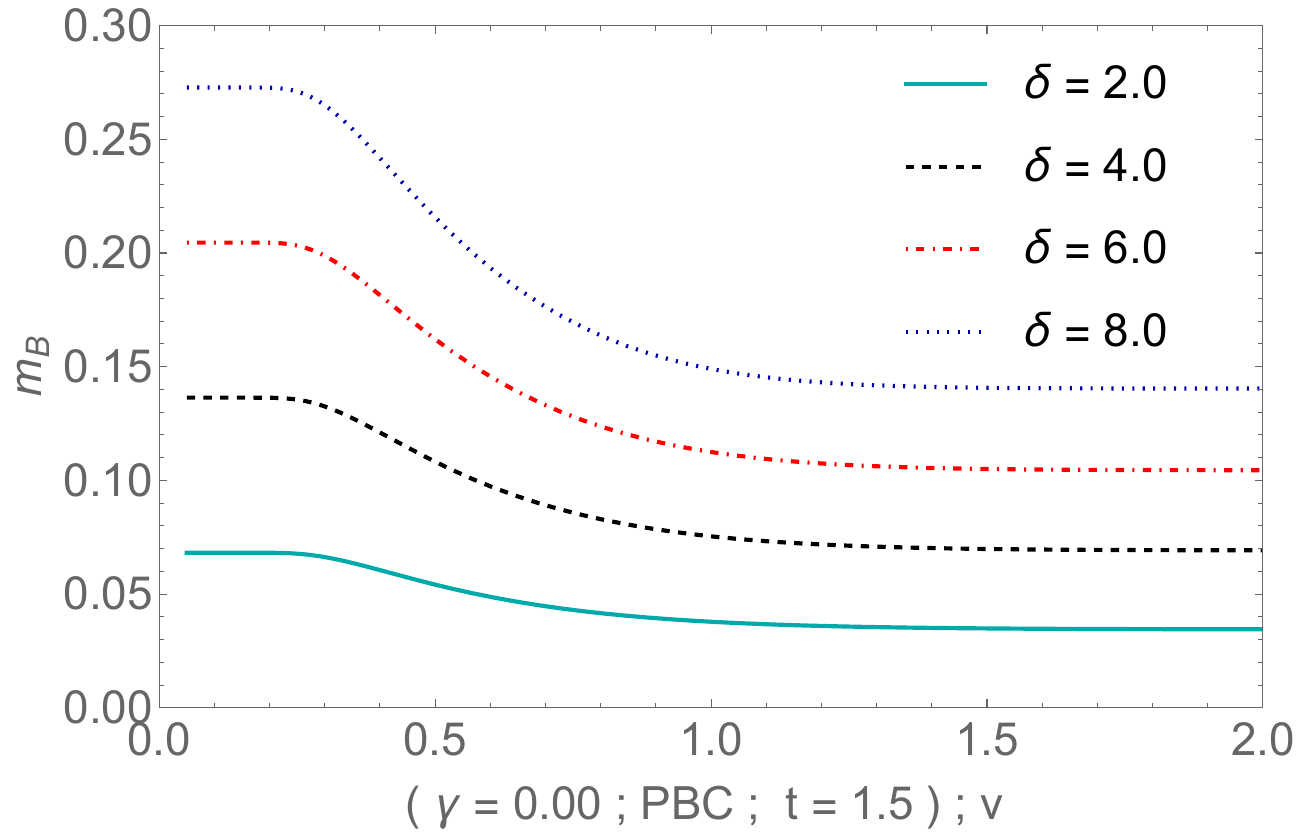} \\
\caption{~Dimensionless magnetization of a Dirac gas as a function of dimensionless volume for several external magnetic fields at fixed reduced temperatures. The left panel corresponds to $t=0.5$, and the right panel to 
$t=1.5$. Boundary conditions are ABC (top), QBC (middle), and PBC (bottom).}
\label{Fig3}
\end{figure}

In Fig.~6, we confirm the catalytic role of the magnetic field in the fermion system. Moreover, no significant differences arise between periodic, antiperiodic, and quasi-periodic boundary conditions at the extremum of the continuum strip of the considered volume (both for small system sizes and for larger volumes, such as in the bulk regime). In general, the PBC and ABC cases exhibit monotonic behavior, whereas the QBC case does not. For twisted boundary conditions with the parameter $\alpha_j = 0.40$ applied in all three spatial directions, for instance, we observe a global minimum.
\begin{figure}
\centering
\includegraphics[{width=6.45cm}]{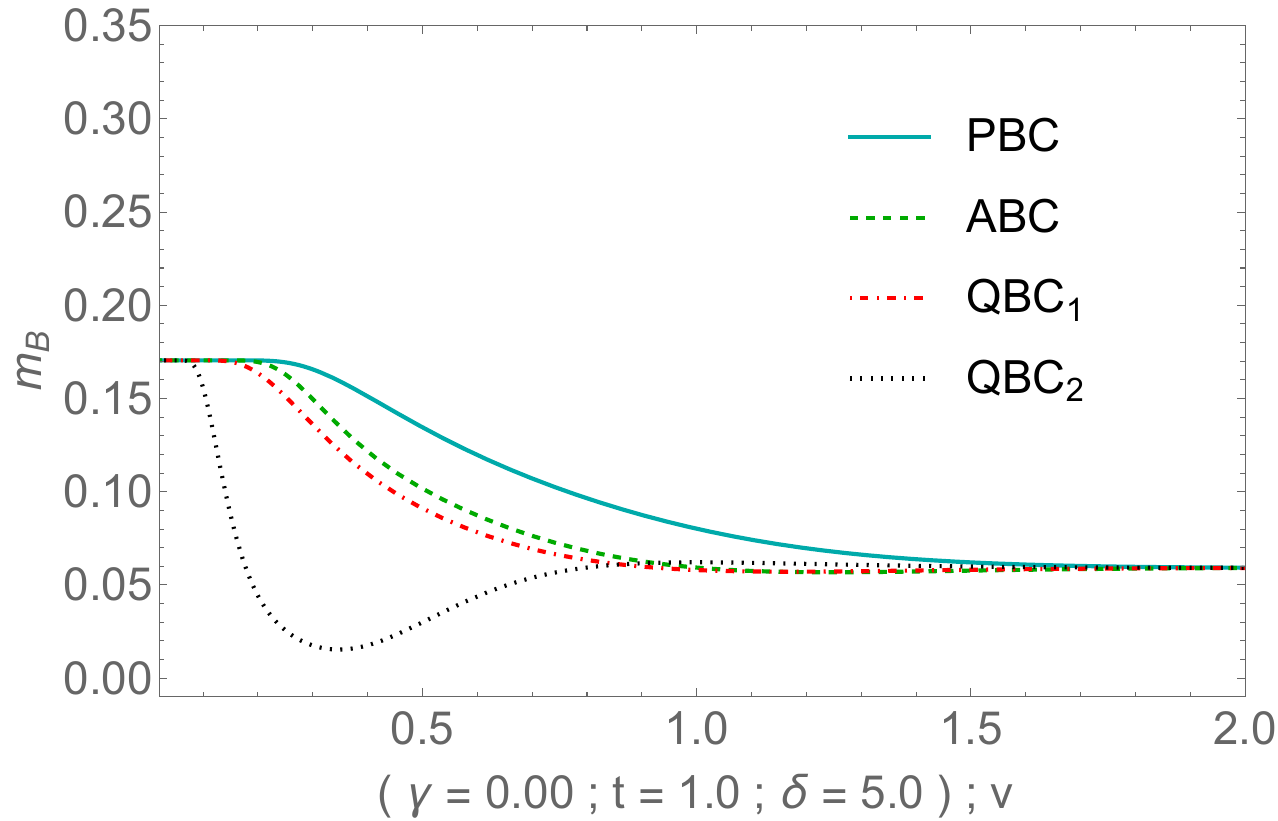}
\includegraphics[{width=6.45cm}]{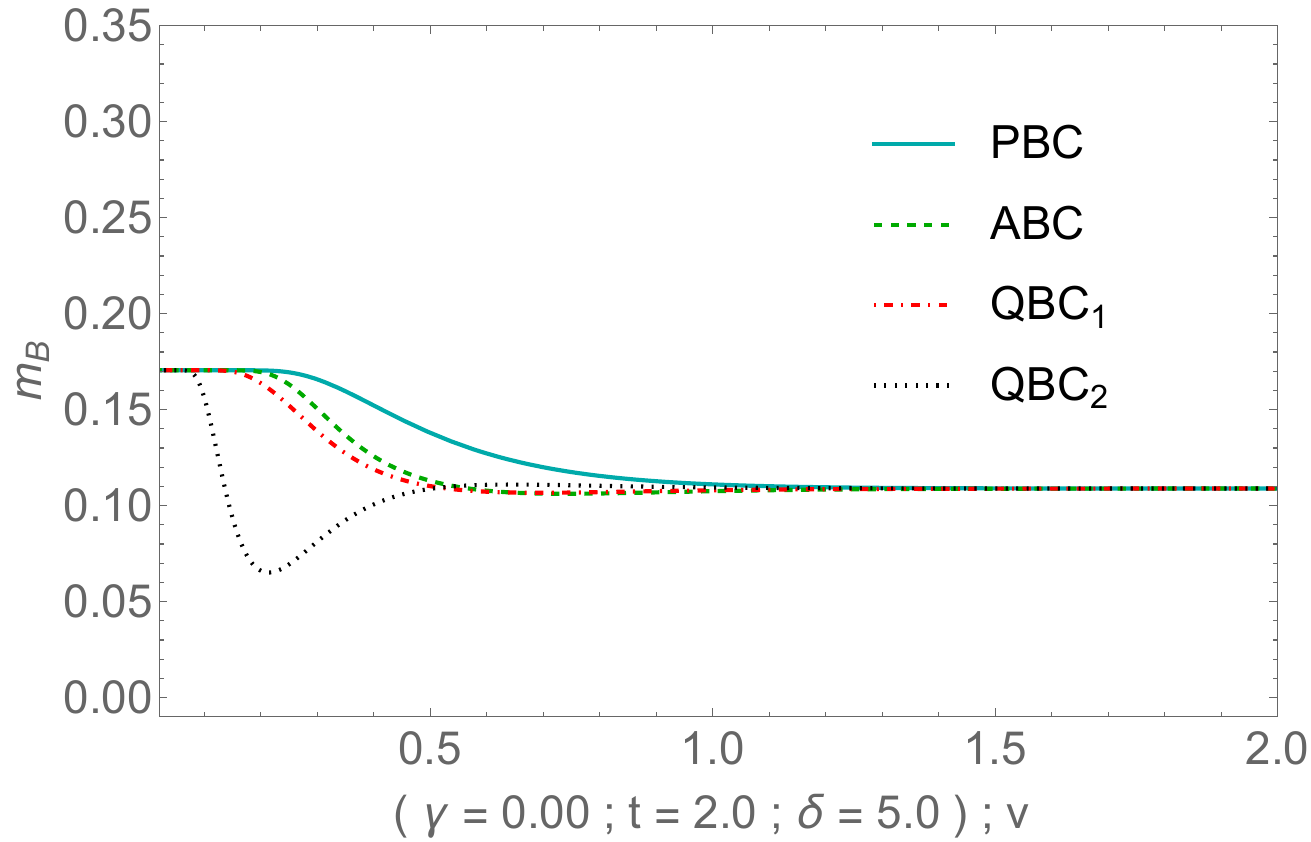} \\
\includegraphics[{width=6.45cm}]{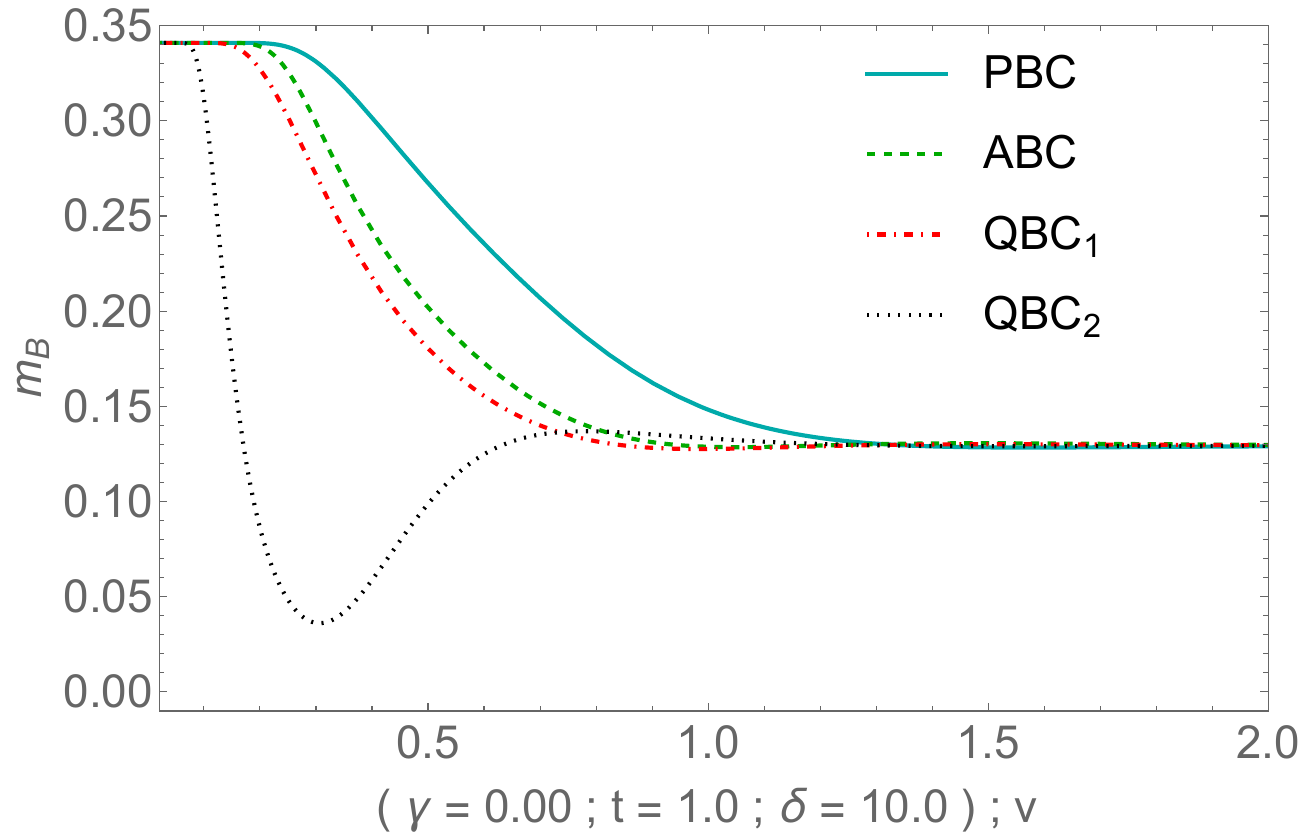}
\includegraphics[{width=6.45cm}]{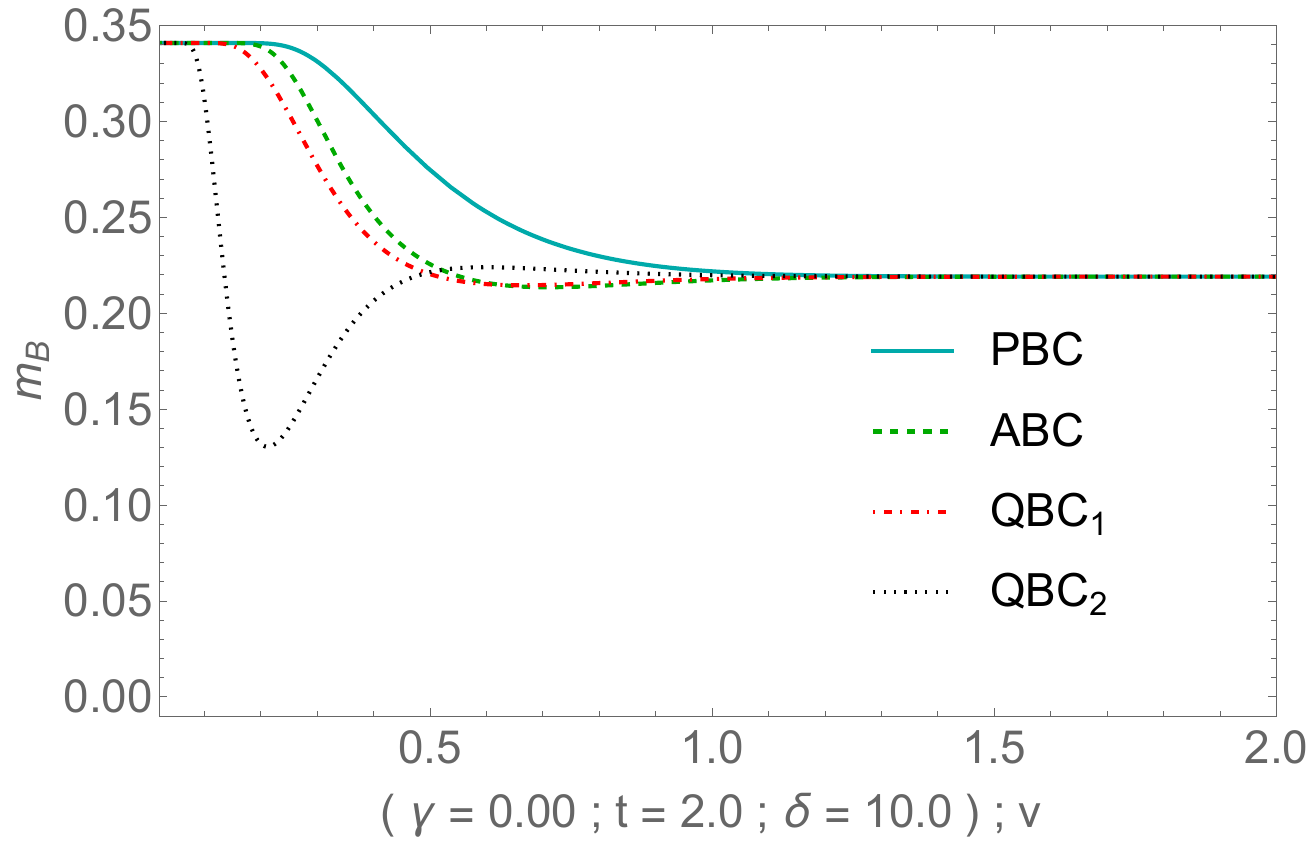} \\
\caption{~Dimensionless magnetization of a Dirac gas as a function of the reduced volume at fixed dimensionless magnetic fields, and reduced temperatures. We consider several types of boundary conditions: ABC ($\alpha_j = 1$), $\mathrm{QBC}_{1}$ ($\alpha_j = 0.70$), $\mathrm{QBC}_{2}$ ($\alpha_j = 0.40$) and PBC ($\alpha_j = 0$), for $j=1,2,3$.}
\label{Fig4}
\end{figure}

In Fig.~7, we plot the behavior of the system for continuous values of the parameter $\alpha_j$. For periodic ($\alpha_j = 0$) and antiperiodic ($\alpha_j = 1$) boundary conditions, the system remains paramagnetic. Nevertheless, in the QBC case, diamagnetic behavior can emerge, particularly at lower temperatures and smaller volumes under weak magnetic fields. However, when the same variables are considered with an edge length twice that of the original box, the system undergoes a transition from the diamagnetic to the paramagnetic phase.
\begin{figure}
\centering
\includegraphics[{width=6.45cm}]{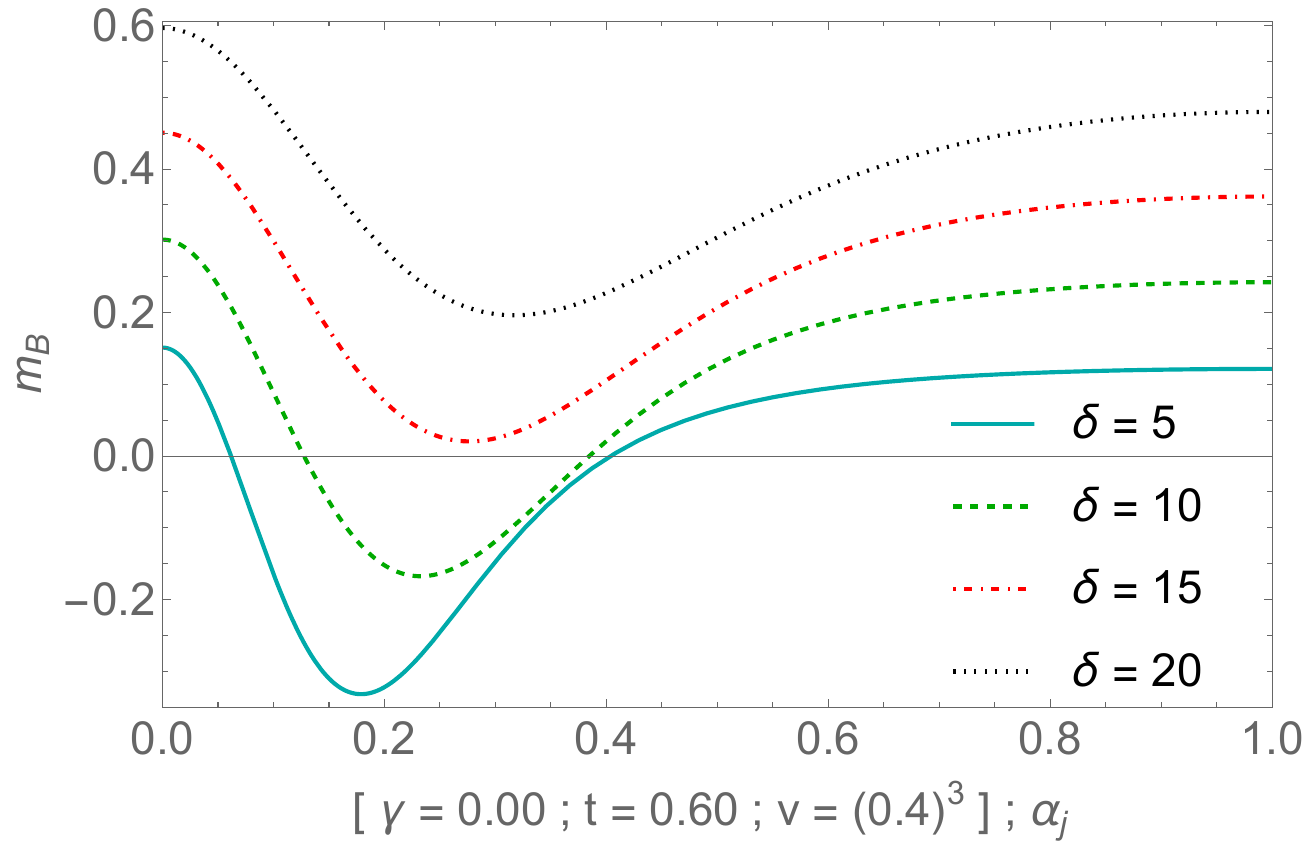}
\includegraphics[{width=6.45cm}]{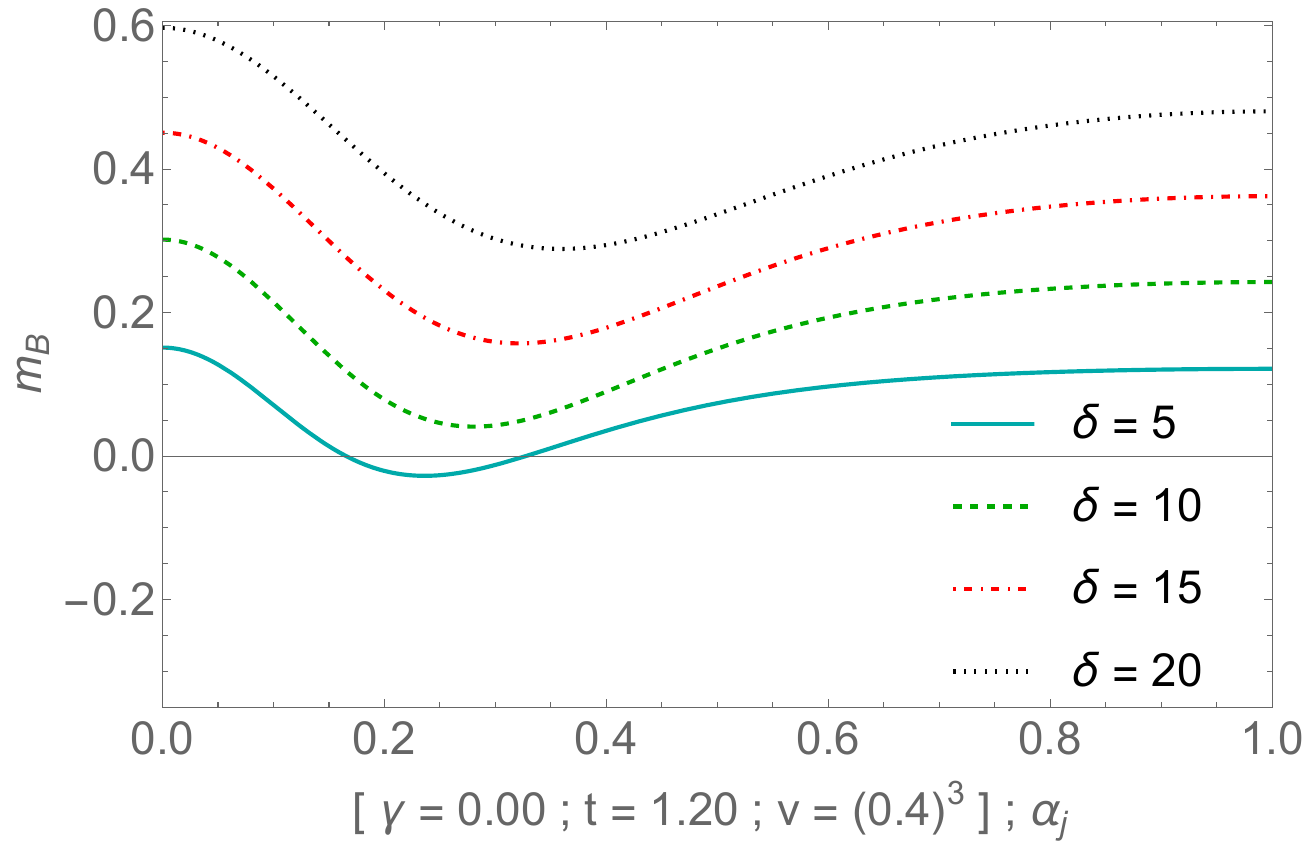} \\
\includegraphics[{width=6.45cm}]{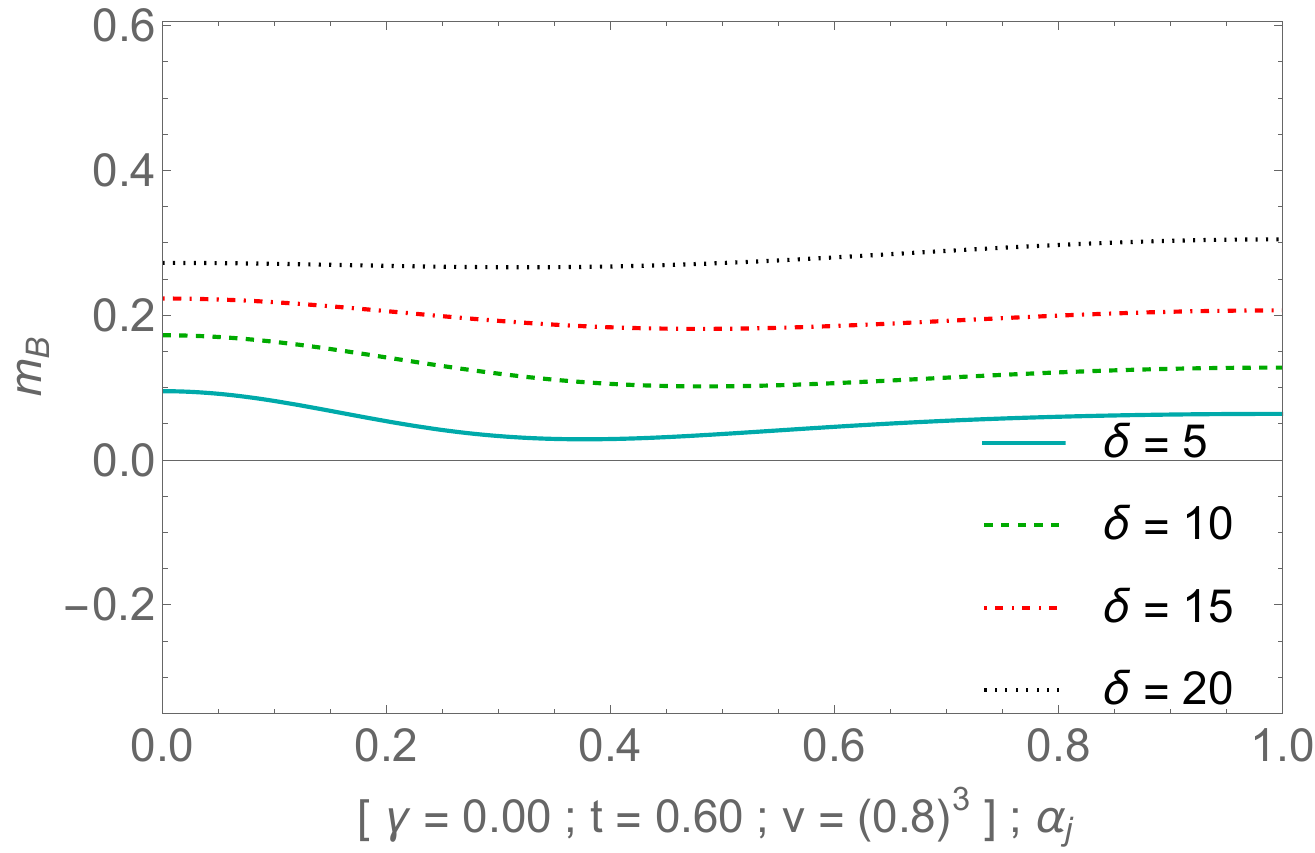}
\includegraphics[{width=6.45cm}]{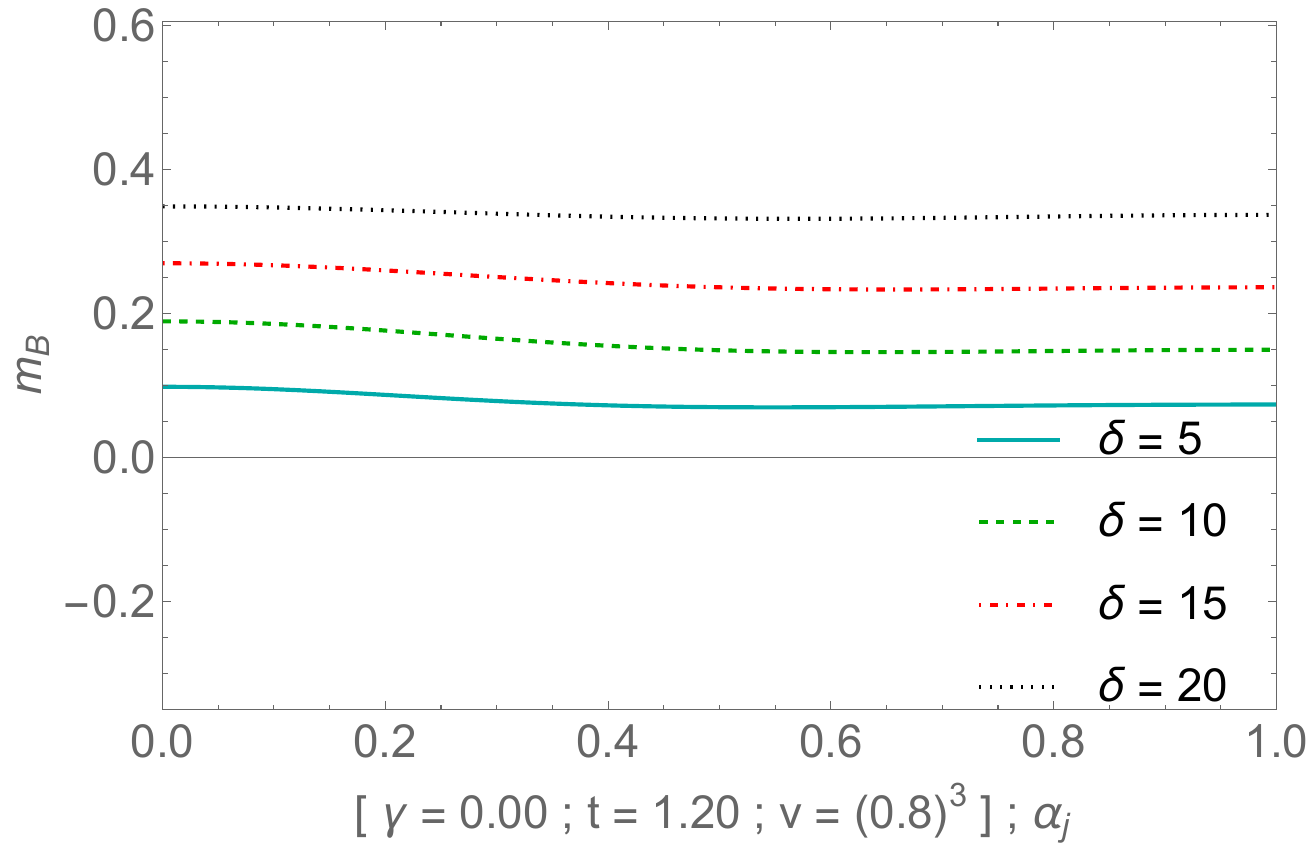} \\
\caption{~Dimensionless magnetization of a Dirac gas as a function of the boundary parameter $\alpha_j$, at fixed dimensionless magnetic field, reduced volume, and reduced temperature. We consider several types of boundary conditions: ABC ($\alpha_j = 1$), QBC ($0 < \alpha_j < 1$), and PBC ($\alpha_j = 0$), for $j = 1, 2, 3$.}
\label{Fig5}
\end{figure}
\section{Comments and conclusions}

We have presented an analytical expression for the effective Lagrangian of a fermionic system under an external magnetic field defined on a toroidal topology. The boundary conditions, represented by the contour parameters $\alpha_j$, are consistent with the transition functions $\varphi_{y} = (\pi\alpha_{y}/e)$ and $\varphi_{z} = (\pi\alpha_{z}/e)$. However, since the transition function in the $x$-direction depended on $y$, we employ the approximation $\varphi_{x} \approx \pi\alpha_{x}/e$.

We considered three types of boundary conditions: antiperiodic, quasiperiodic, and periodic. As an immediate application of the quantum dynamics investigated in this paper, we obtained the magnetization behavior of the system. We found that the chemical potential has only a minor influence on the system behavior.

In the PBC and ABC cases, the magnetization exhibited a paramagnetic nature, whereas in the QBC case, for small volumes, we observed diamagnetic behavior under weak magnetic fields, as shown in Fig.~7. When the volume or magnetic field is increased, this diamagnetic phase undergoes a transition to a paramagnetic phase. We hope to continue studying effective Lagrangians in other external backgrounds in future investigations.

\end{document}